\documentclass[conference]{IEEEtran}
\usepackage{graphicx}
\usepackage{booktabs} 
\usepackage{amsfonts}
\usepackage{pgfplots}
\usetikzlibrary{patterns}
\usepackage{amsmath, amsthm, amssymb}
\usepackage{comment}
\usepackage{capt-of}
\newtheorem{prop}{Proposition} 
\usepackage{amsmath}
\usepackage{algorithm} 
\usepackage{algpseudocode}
\usepackage{subfigure}
\pgfplotsset{compat=newest}

\begin{document}
\title{A Comparative Study of Image Disguising Methods for Confidential Outsourced Learning}

\author{
{\rm Sagar Sharma}\\
Bytedance\\
Seattle, WA\\
sagar.sharma@bytedance.com
\and
{\rm Yuechun Gu, Keke Chen}\\
Trustworthy and Intelligent Computing Lab\\
Marquette University, Milwaukee WI\\
\{ethan.gu, keke.chen\}@marquette.edu
} 

\maketitle
\subsection*{Abstract}
Large training data and expensive model tweaking are standard features of deep learning for images. As a result, data owners often utilize cloud resources to develop large-scale complex models, which raises privacy concerns. Existing solutions are either too expensive to be practical or do not sufficiently protect the confidentiality of data and models. In this paper, we study and compare novel \emph{image disguising} mechanisms, DisguisedNets and InstaHide, aiming to achieve a better trade-off among the level of protection for outsourced DNN model training, the expenses, and the utility of data. DisguisedNets are novel combinations of image blocktization, block-level random permutation, and two block-level secure transformations: random multidimensional projection (RMT) and AES pixel-level encryption (AES).InstaHide is an image mixup and random pixel flipping technique \cite{huang20}. We have analyzed and evaluated them under a multi-level threat model. RMT provides a better security guarantee than InstaHide, under the Level-1 adversarial knowledge with well-preserved model quality.
In contrast, AES provides a security guarantee under the Level-2 adversarial knowledge, but it may affect model quality more. The unique features of image disguising also help us to protect models from model-targeted attacks. We have done an extensive experimental evaluation to understand how these methods work in different settings for different datasets. 

\section{Introduction}
Deep Neural Networks (DNN) have shown impressive performance across diverse domains such as image classification, natural language processing, speech recognition, and recommendation systems. However, DNN training is resource-intensive and time-consuming, requiring large training data, careful model architecture selection, and exhaustive model parameter tweaking. As a result, data owners or model developers often utilize multiple cloud GPUs or online model training services, such as Google Colab, to lower their costs. 

Despite its popularity, outsourcing DNN learning to the cloud raises privacy and security concerns about the sensitive training data and trained models \cite{sharma18ic,duncan12}.
On the one hand, cloud users cannot verifiably prevent the cloud provider from getting access to their data. In practice, using public clouds often means fully trusting your cloud provider. On the other hand, public cloud providers are not immune to security attacks, which may lead to data breaches through insider  \cite{chen10,duncan12} and external attacks \cite{mansfield15,unger15}. Additionally, membership inference attacks \cite{shokri16}, model inversion attacks \cite{fredrikson14}, and adversarial example exploration \cite{chakraborty18,raff19} can be applied to models directly to explore the training examples in DNN learning. Therefore, data and models in training, testing, and transferring between the cloud and the client are seriously threatened.

There have been a few efforts trying to address this critical issue. However, all of them are not satisfactory.
\begin{itemize} 
	\item \emph{Encrypted data and models.} The first approach is to train encrypted DNN models over encrypted data. However, due to the large training data and expensive training process in deep learning, cryptographic model training approaches are too expensive to be practical. A recent study on training small-scale neural networks  ~\cite{mohassel17} (e.g., just two layers with a maximum of 128 neurons per layer)  has shown astonishingly high communication, computation, and storage costs. As a result, cryptographic approaches are often limited to training small models \cite{mohassel17,sharma19} or securely applying trained DNNs for prediction \cite{xie14,rathee20,huang20}. 

	\item \emph{Federated learning.} Another possible solution is to partition the dataset and the learning task into sensitive and non-sensitive partitions and use cloud-client federated learning \cite{kairouz19}. The non-sensitive portion, assuming it is much larger than the sensitive one, is exported to and processed by the cloud. Correspondingly, the learning process is partitioned and distributed between the cloud and the client, ensuring that the intermediate information exchanged between the two parties does not breach privacy. Collaborative deep learning framework enhanced with differential privacy \cite{reza15,abadi16} may be tweaked into such a partition-based setting. However, reported attacks \cite{hitaj17}  allow an adversarial collaborator (e.g., a compromised cloud in the cloud-client scenario) to generate images resembling the sensitive classes owned by the victim parties (the trusted data owner).

\item \emph{Trusted execution environments.} Hardware-assisted trusted execution environments (TEEs), such as Intel SGX, can also be applied to deep learning in the cloud. The idea is to create a secure enclave in the specific memory area (enclave page cache (EPC)) so that no other process/thread can access the content in the enclave. Memory pages are also automatically encrypted when they are swapped to the disk. However, recent studies on side-channel attacks \cite{fei21} make this approach challenging to develop and deploy. Attackers can peek or infer the plaintext content inside the secure enclave via side channels, such as page fault interrupts and cache loading. 

Furthermore, to work with GPUs, costly cryptographic approaches have been applied to achieve partial data confidentiality in transferring data between CPU and GPU \cite{tramer18,ng21}, which does not meet the performance requirement for training large DNNs or with large training data. GPU manufacturers may develop TEEs for GPU \footnote{Nvidia has announced a GPU TEE in their Hopper architecture.} However, it's to be tested to determine how secure it is in terms of side-channel attacks. 

\end{itemize}

Researchers have also explored the application of differential privacy (DP) \cite{dwork06} in distributed (federated) learning scenarios \cite{reza15} or a trusted central training server  \cite{abadi16}. However, DP works for the setting of sharing data and models without breaching individual training examples' privacy. It does not meet the need for data and model confidentiality.

\textbf{Scope and contributions.} 
This study compares two different image disguising approaches:  InstaHide \cite{huang20} and our recently developed DisguisedNets. These image disguising mechanisms protect the training data and also possibly the learned models by casting training data into a confidential transformed space where powerful DNN models can still learn features and patterns distinguishing image classes and leverage the power of GPUs in the cloud. The intuition is twofold. (1) Apply appropriate transformations and data protection mechanisms so that the disguised images cannot be effectively reconstructed and re-link to the original images. (2) Meanwhile, powerful deep learning techniques can still pick up the unique topological and geometric features preserved in the transformed space to distinguish the originally defined classes of images in the transformed space. By doing so, the tie between the original training data and the learned model in the transformed space is broken, which also disables any model-based exploration \cite{chakraborty18,raff19,shokri16,fredrikson14}. In the end, we can approximately preserve the \emph{image distinguishability} for the target classification task while minimizing the \emph{recoverability} of individual images.


There are several unique contributions. 
\begin{enumerate}
\item We have designed two image disguising mechanisms: AES-based (AES) and random-projection-based (RMT) for image-based DNN learning to preserve training data and model confidentiality in outsourced training. The goal is to study and achieve a good balance between the utility of disguised images and the level of confidentiality protection. 
\item We have carefully analyzed the potential attacks under the outsourced deep learning settings and the resilience of disguising mechanisms to attacks on data and model confidentiality. So that users can choose the corresponding method under their preferred threat model.  
\item 
We have conducted extensive experimental evaluations on public datasets to show the trade-offs of different disguising schemes and related parameter settings between data utility and their resilience to attacks. We also show how the disguising methods work to protect models from model-targeted attacks and 
\end{enumerate}


\section{Related Work}
Sensitive deep learning assets may include training data, models, and online testing data. Protection methods may target the model training phase or the application (i.e., inference) phase when both phases can be exported to the public cloud. However, typically, the computational complexity and the demand on resources of model training is far more than model application. Thus, many studies have focused on the application phase, e.g., a cloud-based model inference service hosting the trained model and making predictions for a user-provided image \cite{gilad16,rathee20,huang22}. 

In contrast, due to the high computational complexity of training algorithms and the large size of training data, there is no practical cryptographic approach, e.g., homomorphic encryption or secure multi-party computing based approaches for protecting model training. A few recent studies in this direction have shown prohibitively high costs even for a small neural network model \cite{mohassel17,sharma19}. Trusted execution environments (TEE) with masked GPU operations are applied to speed up training \cite{ng21,tramer18}. However, no TEE-based deep learning method has addressed severe side-channel attacks \cite{fei21}. 

\begin{figure}[h]
  \centering
    \includegraphics[width=.60\linewidth]{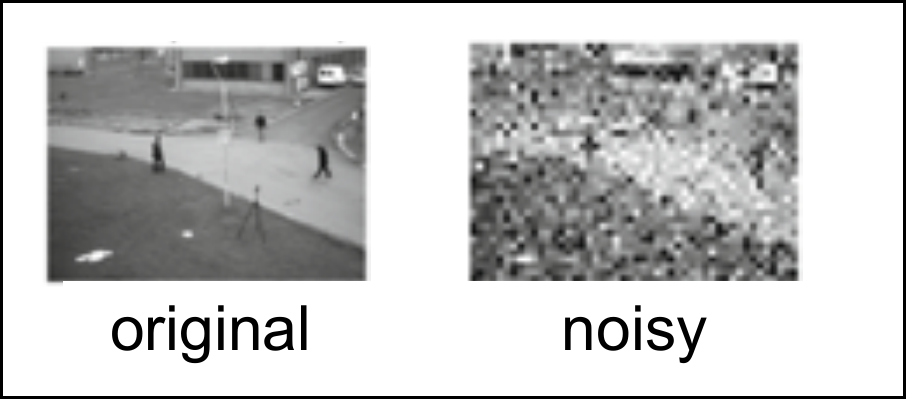}\\
    \small (a)\\
    \includegraphics[width=.60\linewidth]{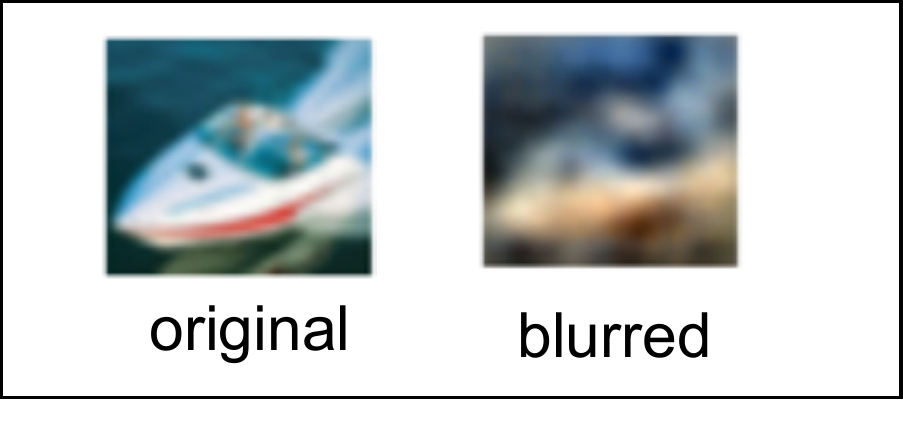}\\
    \small (b)\\
    \includegraphics[width=.60\linewidth]{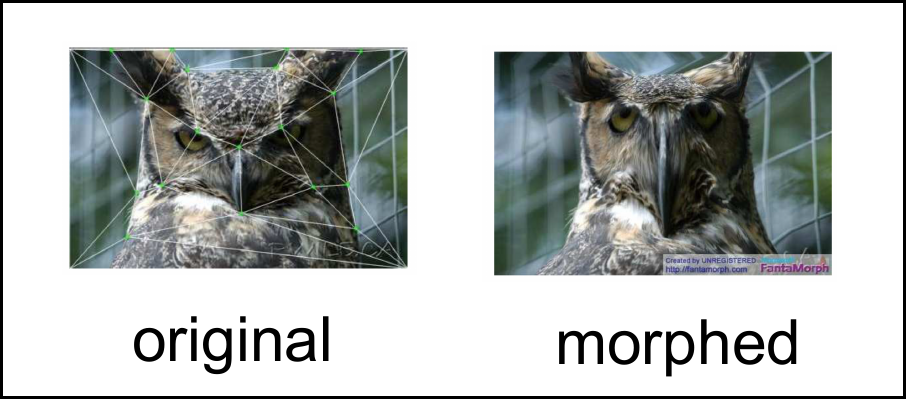}\\
    \small (c)\\
  \caption{(a) Differentially private noise addition to images ~\cite{fan18}; (b) The reconstructed blurred images in PrivyNet ~\cite{li17}; (c) The morphed images ~\cite{lee96_morph}} 
  \label{fig:existing_app}
\end{figure}

Researchers have looked at protection methods for image training data in the outsourcing context. Noise addition \cite{fan18}, image blurring \cite{li17}, and morphing \cite{lee96_morph} are weak as the visible features of the images are still perceivable and understandable. Such transformations do not defeat simple visual re-identification. Figure \ref{fig:existing_app} shows how easy it is to visually re-identify the content in the original images by observing the transformed ones with the mentioned techniques. 

A recent method InstaHide \cite{huang20} applies the idea of mixing up images \cite{zhang18} in a training set with public images with linear combination to obfuscate the content,  along with randomized signs of pixel values for further protection. However, it is vulnerable to image reconstruction attacks \cite{carlini21} that need only to know the disguised images and public image sets (i.e., the Level-1 adversarial knowledge, as we will discuss). 

\begin{table*} [h!]	\centering
	\small
	\caption{Related work on training phase protection.}\label{tab:related_work}
	\begin{tabular}{|p{2cm}|p{4.5cm}|p{6.0cm}|p{4.5cm}|} \hline
		{Sample Related Work}&Method&Weaknesses&Strengths\\ \hline 
		Mohassel et al. \cite{mohassel17} & Train DNN models over masked or encrypted data with cryptographic protocols. & Involve high communication, computation, and storage costs; Require extensive re-design and custom implementation of the DNN architecture; Involve iterative interactions between the data owner and cloud provider. & Provide semantic security.  Model quality is fully preserved. \\ \hline
		Tramer et al. \cite{tramer18} and Ng et al. \cite{ng21} & Use TEEs for confidential CPU operations and masked data for confidential GPU operations & side channel attacks are an unaddressed concern & More efficient than cryptographic protocols, but GPU operations on masked data are still expensive\\ \hline
		Abadi et al. \cite{abadi16} and Shokri et al. \cite{shokri15} & Apply differential privacy to randomize intermediate gradients & Aim to share trained models, and thus does not preserve model confidentiality; result in significant drop in model quality; vulnerable to generative adversarial network attacks. & Preserve individuals' privacy. \\ \hline
	Fan \cite{fan18} & Train shallow neural network locally and outsources intermediate representation to the cloud for deeper training. & The intermediate representation of images reveals the visual characteristics of the related images, vulnerable to visual re-identification attacks.  & NA \\ \hline
		Li et al \cite{li17} & Applies differential privacy to hide sensitive pixels in images & Does not hide the global visual characteristic content of images, vulnerable to visual re-identification attacks & NA \\ \hline
		Zhang et al. \cite{zhang18} and Huang et al. \cite{huang20} & Mix-up images from the training set and public domain with random selection and weighing to hide the content of the sensitive image. & Vulnerable to ciphertext-only (Level 1 adversarial knowledge) image re-construction attacks; May expose trained model to wide variety of model-based and membership attacks for the inside-dataset setting.  & Fast training; No changes to the training architecture.\\ \hline
	\end{tabular}
\end{table*}

Thus, on the one end, existing disguising mechanisms are too weak to protect almost nothing. On the other end, if an encryption mechanism, e.g., homomorphic encryption, or a complex cryptographic protocol, is applied, such linking or reconstruction would be impossible. However, the current cryptographic schemes incur extremely high costs in almost all aspects of computation, communication, and storage. Thus, they are impractical for resource-intensive tasks like training a DNN model. Hardware-assisted approaches, such as TEEs, are still under investigation to ensure the expected security properties. Along with all these possible approaches, we aim to explore and develop new image-disguising methods to achieve good balances among costs, data utility, and security guarantees. Table \ref{tab:related_work} summarizes the current work in protecting data and models in the training phase.

\section{Threat Modeling} 
We are concerned with the confidentiality of the sensitive training image data and the DNN models in the \emph{outsourced training phase}. Here, we make some relevant security assumptions for our disguising mechanisms. 1) We consider the cloud provider to be an honest-but-curious adversary, which implies that a curious provider will still honestly deliver desired results to the data owner. However, it may keep a copy of the data and programs it can observe. 2) The adversary can observe the training data, the training process, and the trained models, including the structure of the DNN architecture and parameter settings for training. Thus, they can probe the observed items with methods such as image reconstruction, re-identification, and membership attacks. (3) We do not address evasive attacks and poisoning attacks \cite{chakraborty18,raff19}, where adversaries will tamper with the training data, which can be guarded with training data integrity checking. (5) The client infrastructure and communication channels are secure. 

\textbf{Assets to Be Protected.} Under a certain protection mechanism, we generalize that the training data $D$ is transformed to $f_{key}(D)$, and the model $M=M(D)$ is changed to $g_{key}(D)$. The attacker might want to know whether an image is likely used to train a model, e.g., the membership inference attack \cite{shokri16}, observe training images that may contain sensitive objects, or steal a proprietary training dataset. The attack might also target the model if the model is exposed, e.g., using model-inversion attacks \cite{fredrikson15} to explore the private information of training data. 

\textbf{Adversarial Prior Knowledge.} The adversary may have two levels of prior knowledge. For each level, we may design a disguising technique. 
\begin{itemize}
\item \emph{Level-1:} They may know what the model is used for, e.g., the background application, the distribution of the data, e.g., face images, and the type of disguising technique used, but do not know the disguising parameter setting for a specific dataset that serves as the secret key to the protocol. 
\item \emph{Level-2:} In addition to Level-1 knowledge, they may try to obtain pairs of images and their disguised versions via other attacking channels (not including the ones they are targeting). They hope to use these known pairs to explore various image reconstruction attacks.
\end{itemize}

\textbf{Potential Attacks.} Recent studies have shown that attackers can explore training/testing examples and models, for example, to find adversarial examples misleading the prediction of deep neural networks \cite{chakraborty18, raff19}. Such attacks depend on adversaries' clear understanding of the original image data and the ability to access the developed models freely. Outsourced learning without protection makes these attacks easier to deploy. 

With a protection mechanism on data and models, we consider a fundamental attack: \emph{training image re-identification} that aims at linking the protected images to identifiable original images. We introduce a model-based re-identification test -- DNN examiner, which uses a model trained on the original data to tell whether a protected image is re-identifiable. Note that some related methods \cite{fan18,li17} are not resilient to human visual re-identification, which does not protect confidentiality. Since the image disguising mechanisms break the link between the original training data and the learned models (in the transformed space), the existing model-oriented attacks do not work anymore without successfully breaking the disguising mechanism. Attackers thus depend on \emph{reconstruction attacks}: reverse the disguising mechanism to approximately reconstruct the original images and then try to re-identify the reconstructed images. We use the DNN examiner approach to evaluate how successful a reconstruction attack works in our experiments.

\section{DisguisedNets -- a Novel Image Disguising Mechanism for Outsourced Deep Learning}\label{sec:core}
In the following, we will introduce an image disguising framework that incorporates pixel-block partitioning, random block permutation, and block-wise transformations of images along with noise additions. The premise is that after the dramatic transformation, it is difficult to link the disguised images to the original images, while, unlike pure encryption schemes, it still preserves some essential patterns for distinguishing between classes of images that allow DNN learning methods to capture. This amalgam of multiple transformations provides a sufficiently large parameter space so that the attacks are computationally intractable (Section \ref{sec:security_analysis}) under the Level-1 prior knowledge. 

\begin{figure} [h]
\centering
\includegraphics[width= 0.9\linewidth]{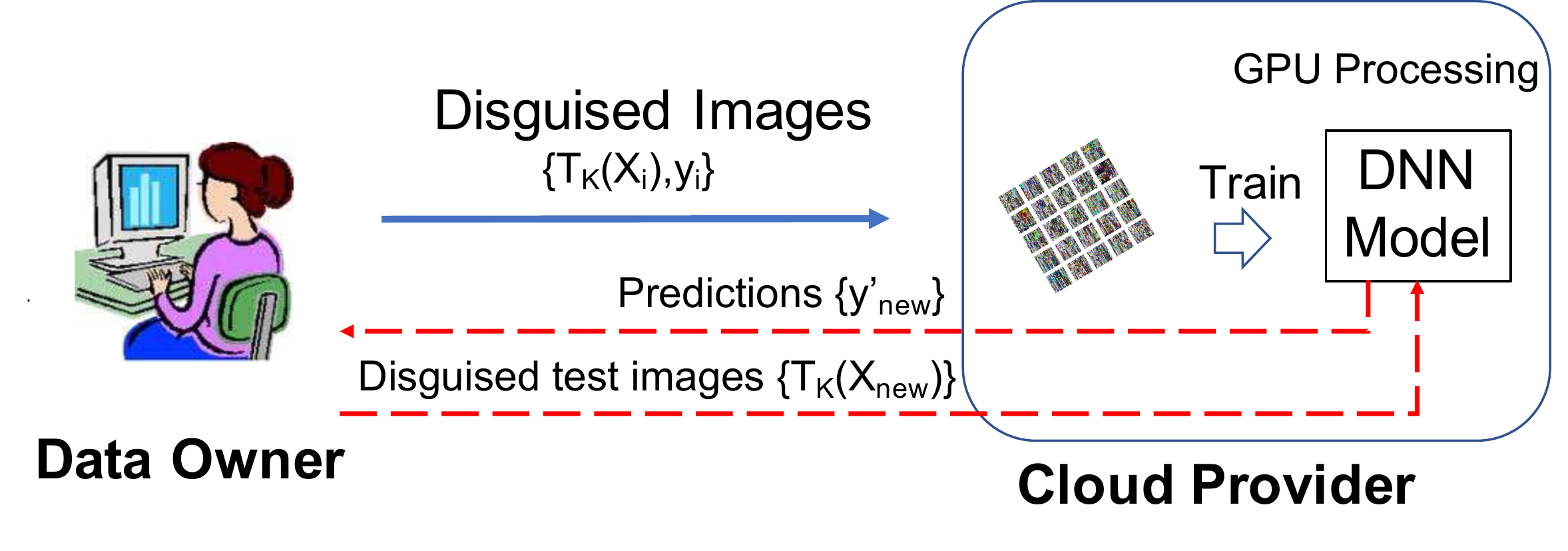}
\caption{Image disguising framework for DNN learning.}
\label{fig:framework}
\end{figure}

Figure ~\ref{fig:framework} depicts the DisguisedNets framework.  A data owner disguises her private images before outsourcing them to the cloud for DNN learning. She can either fully outsource the entire image datasets and the learning procedure to the cloud or selectively retain sensitive images in the cloud-client partitioning setting. She transforms all of her images using one secure transformation key secret to her. Note that this transformation should be at a reasonable cost, practical for a client's infrastructure to process. 

Specifically, assume the data owner owns a set of images for training, notated as pairs $D=\{(X_i, y_i)\}$, where $X_i$ is the image pixel matrix (${l\times m}$ and ${3\times l \times m}$ for grayscale and RGB images respectively) and $y_i$ the corresponding label. We formally define a disguising mechanism as follows. Let the disguising mechanism be a transformation $T_K$, where $K$ is the secret key that depends on the selected perturbation techniques. By applying image disguising, the training data is transformed to $\{(T(X_i), c_i)\}$ with $c_i$ mapped to $0,1 , \dots$ randomly representing the classes $y_i$.  The original model is a function $\mathbf{M}(D)$, which is learned with a DNN learning method $\mathfrak{G}$: $\mathfrak{G}(D) \rightarrow \mathbf{M}(D)$. The image disguising mechanism enables the same learning method $\mathfrak{G}$ to be applied to the transformed data directly without any modification:  $\mathfrak{G}(D_T) \rightarrow \mathbf{M}_T(D_T)$. For any new data $X_{new}$, the model application (or inference) is defined as $\mathbf{M}_T(T(X_{new}))$, i.e., the new data transformed with the same key. 
To make such a transformation method practical for modeling, i.e., a model trained with transformed data still working satisfactorily, a user may expect the error of modeling is not far away from the original model's. Thus, a utility-preserving mechanism should have 
\[
|Err(\mathbf{M}_T) - Err(\mathbf{M})| < \delta
\]
where $\delta$ is the level of model quality degradation acceptable to the user. While for a specific DNN modeling method and a specific dataset, it's difficult  to theoretically justify what this gap will be, one can always directly evaluate the model quality to check whether it is acceptable for the application. We have empirically evaluated the $\delta$ levels for different mechanisms, datasets, and a few popular DNN modeling architectures in experiments.

\subsection{Pixel-Block Partitioning and Block-based Random Permutation}\label{subsec:perm} 
In this section, we present one way of image transformation: image block permutation, that will be combined with other mechanisms later. 

An image $X_{l \times m}$ is first partitioned into $t$ blocks of uniform size $ r\times s$. If we label the blocks sequentially as $v= <1,2,3,4, \dots, t>$, a pseudorandom permutation of the image, $T_{\pi}(X)$, shuffles the blocks and reassemble the corresponding image accordingly. Block-based permutation preserves the in-block information and the relative positions of related blocks. Thus, we understand it preserves a great amount of  information for effective modeling. However, while the permutation may break the global patterns of the images and achieve good visual privacy already, the between-block characteristics such as boundaries, color, content shape, and texture of the original neighboring blocks may provide clues for adversaries to recover the original image -- imagine the jigsaw puzzle!  For large $t$, such attacks can be time-consuming due to the vague similarity between block boundaries. However, with the prior knowledge: a pair of original image and its block-permuted image, it's not difficult to solve such a jigsaw puzzle. Thus, we use this as an auxiliary step enhancing other steps in the disguising framework. 

\subsection{Pixel-Block Transformations}\label{subsubsec:tech_gdp}
Next, we establish pixel-block-level protection mechanisms that aim to preserve the data utility for DNN modeling and further increase the resilience to attacks. We consider two candidate mechanisms: random projection and encryption schemes, and discuss their characteristics. 
Specifically, when an image is partitioned into $t$ pixel blocks for random permutation, we get a list of $t$ parameters $\{K_i, i=1\dots t\}$, one for the pixel-block at the same position across the whole dataset. We name the specific position of the pixel block in the image \emph{the pixel-block position}. The list of parameters acts as a secret key and will be shared, together with the permutation key, by each image in the dataset. The purpose of this setting is to maximize the preservation of distinguishable patterns between image classes -- i.e., a pair of similar image patterns (blocks) can still be transformed to another pair of (likely) similar ones after applying the disguising mechanism.

\begin{figure}[h]
  \centering
    \includegraphics[width=.90\linewidth]{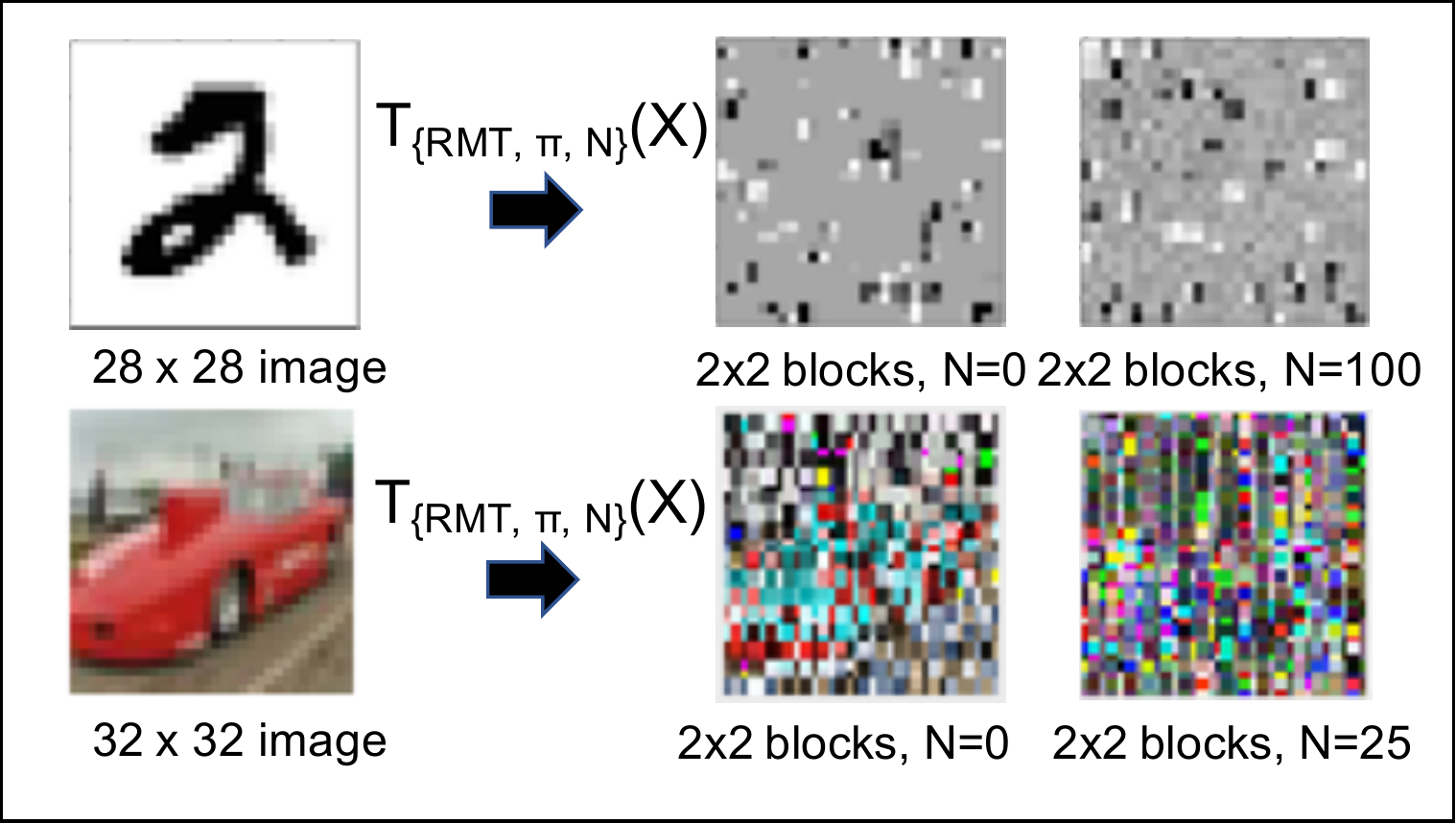}
   \caption{Block-wise RMT+Noise on MNIST and CIFAR-10 images. }
  \label{fig:disguises}
\end{figure}

\textbf{Randomized Multidimensional Transformation (RMT).} For an image represented as a pixel matrix $X$, a general linear transformation can be defined as $G(X) = R(X+\Delta)$, where $R_{ m \times m}$ is a random orthogonal matrix generated following the Haar distribution \cite{gallier00}, or a random invertible matrix, e.g., a random projection matrix \cite{vempala05}, and $\Delta$ is an optional noise matrix. We call this method the randomized multidimensional transformation. When an image is partitioned into $t$ blocks for random permutation, we prepare a list of random matrices $\{R_i, i=1..t\}$, one for each image-block position and share this list for each image. Such transformation is known to preserve (or approximately preserve by random projection) the Euclidean distance between columns of the matrix $X$. For real application, we may arrange the pixel blocks accordingly to form the column of $X$. For example, a 4x4 pixel matrix can be partitioned into 4 2x2 block to preserve the smaller block-level similarity with RMT. Figure ~\ref{fig:disguises} shows the effects of RMT on MNIST and CIFAR-10 datasets. 

\begin{algorithm}
\caption{DN\_RMT (X, t, Key)}\label{alg:rmt}
\begin{algorithmic}[1]
\Require X: image of size $l \times m$; t: number of blocks; Key = \{permutation\_key, transformation\_matrices, noise\_level $\in$ [0, N]\} 

\State r, s $\leftarrow$ compute image block size with $l\times m$ and $t$;
\State Partition image $X_{l \times m}$ into blocks $X_1, X_2,\ldots, X_t$;
\State Shuffle the image blocks pseudorandomly with permutation\_key 
\For{ each block $i, i = 1\dots t$}
	\State $\Delta_i \leftarrow$  Generate random matrix with elements from the uniform distribution in [0,N];
	\State use the transformation matrix at the position $i$: $R_i$;
	\State $Y_i \leftarrow  R_i(X_i + \Delta_i)$;
\EndFor
\State Re-assemble $\{Y_i\}$ to make the transformed image $Y$ and return $Y$;
\end{algorithmic}
\end{algorithm}

\begin{figure}[h]
  \centering
  \includegraphics[width=.60\linewidth]{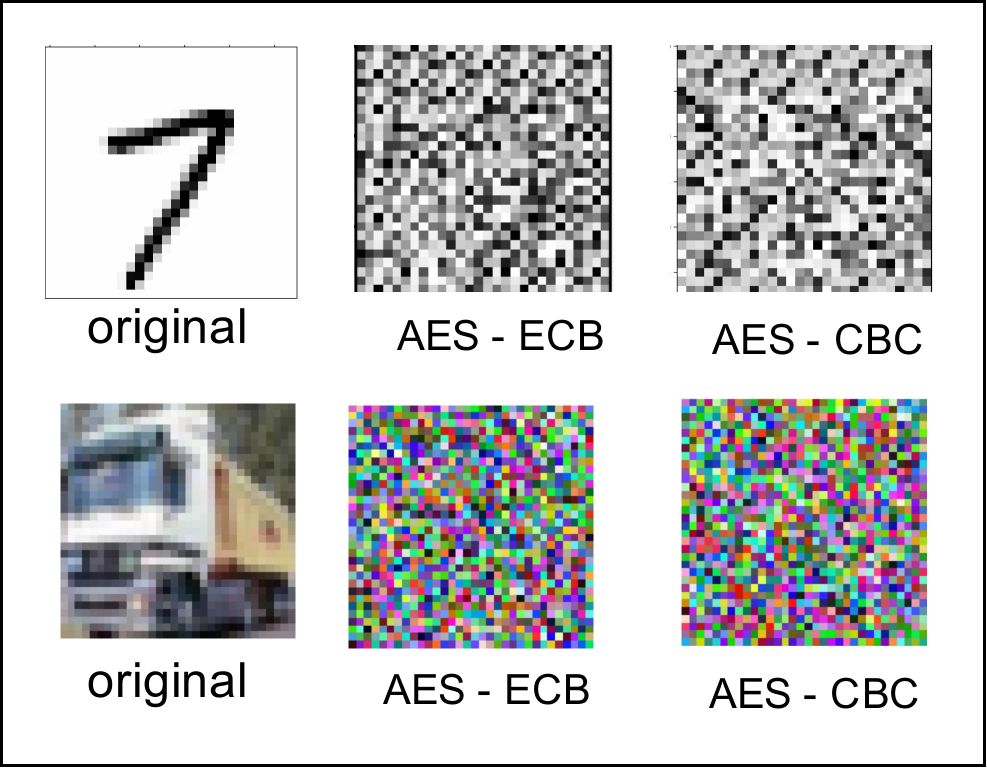}
   \caption{Pixel-block based AES encryption of MNIST and CIFAR-10 images. }
  
  \label{fig:aes_enc}
\end{figure}

\textbf{AES Block Transformation (AES).} The existing AES encryption schemes typically use 128-bit encryption keys, which encode every 16-byte data block sequentially. If we use AES for pixel-block encryption, assuming each pixel is stored in one byte, 16 original pixels are mapped to 16 encrypted bytes (pixels), and a whole pixel block is encoded to 16-byte units. Putting all encrypted pixel blocks together, we get a disguised image. For clear presentation, when we talk about AES encryption block, i.e., 16 bytes for a 128-bit encryption key, we use the 16-byte ``encryption unit'', which are different from ``pixel blocks'' we have been using previously in our image disguising framework. Figure \ref{fig:aes_enc} shows some example AES transformations on images.

\begin{figure}[h]
  \centering
  \includegraphics[width=.80\linewidth]{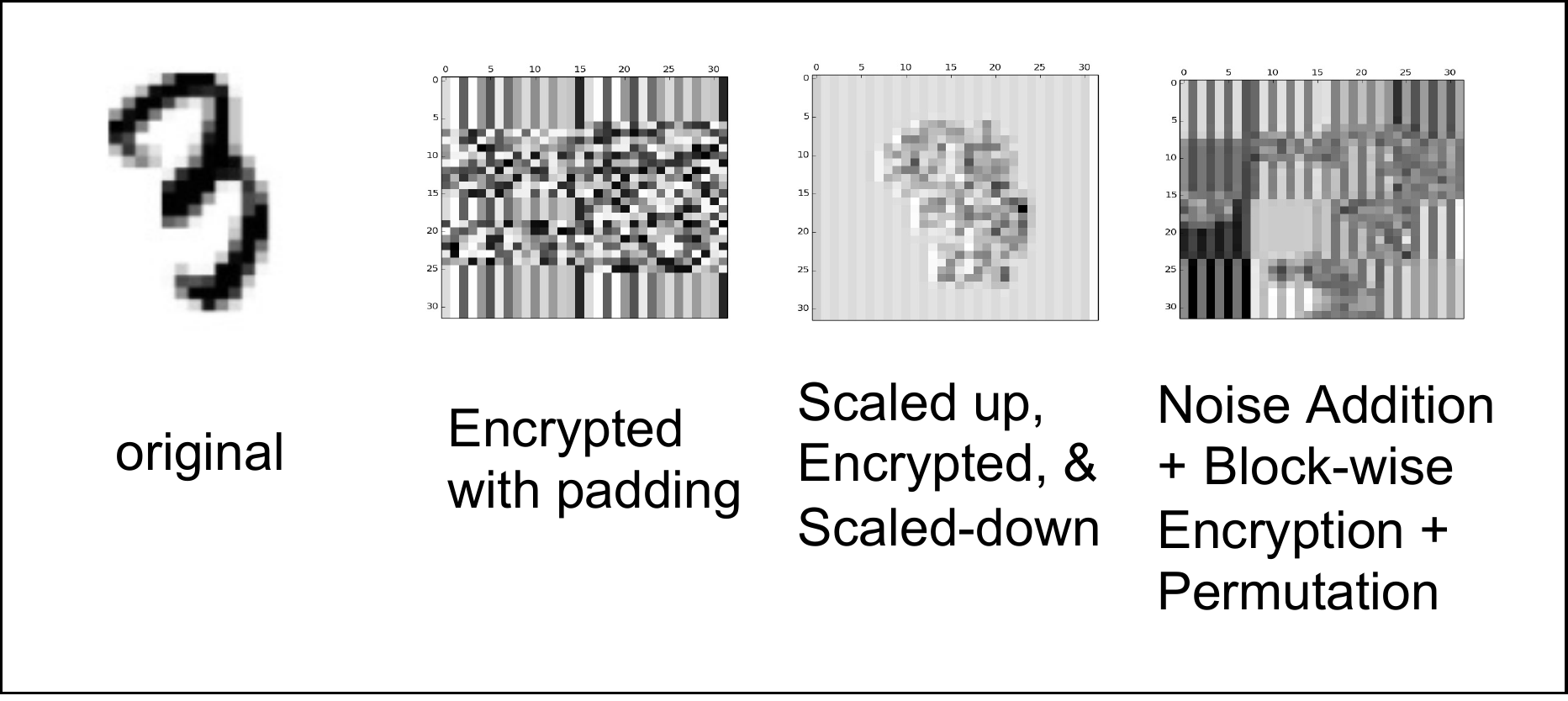}
  \caption{AES-ECB encryption of MNIST image with different strategy. }
  \label{fig:aes_variation}
\end{figure}

We consider two AES modes in our design. (1) We observe that with the AES Cipher Block Chaining (CBC) mode, any pixel-level change in the pixel block between two images will result in different encoding results for most 16-byte AES blocks in this pixel block position, making it not ideal for our purpose.
(2) Then, we turn to the AES Electronic Code Book (ECB) mode that can be considered as a fixed mapping function between 16-byte original data to 16-byte encrypted data. Different from CBC, the neighboring 16-byte blocks do not affect the encoding of the current block. This matches our requirement of data utility preservation, e.g., to preserve the block-level distinguishable patterns after the transformation. 

To preserve more information, based on the intuition of smaller blocks preserving more inter-pixel-block information, we can also use unit sizes smaller than the regular size, which is 16 pixels for 128-bit ECB. The method is to scale up the image first, e.g., from 32x32 to 256x256 (where each pixel is duplicated eight times), then encrypt it by 16-pixel units, and finally scale down to the size 32x32. Please refer to Figure \ref{fig:aes_variation} for the detailed example. It's equivalent to encrypt 2-pixel units in the original 32x32 images. We found by reducing the block size, the model quality can be improved with the cost of lower attack resilience to the Level-2-knowledge-based attack.

\begin{algorithm}
\caption{DN\_AES (X, k, t, Key)}\label{alg:aes}
\begin{algorithmic}[1]
\Require X: image of size $l \times m$; k: scale-up factor; t: number of blocks; Key = \{permutation\_key, AES\_keys, p = probability of salt-pepper noise \} 
\Ensure the selection of k and t results in image blocks that can be further partitioned to 4x4 pixel patches; 
\State $X_{lk \times mk} \leftarrow$ scaled up the image;
\State r, s $\leftarrow$ compute image block size with $lk \times mk$ and $t$;
\State Partition image $X_{l \times m}$ into blocks $X_1, X_2,\ldots, X_t$;
\State Shuffle the image blocks pseudorandomly with permutation\_key;
\For{each block $i, i=1\dots t$}
	\State $Y_i \leftarrow $ for each pixel in block $X_i$, with the probability $p$, it's randomly turned to white or block pixel (salt-pepper noise);
	\State $E(Y_i) \leftarrow $  with the AES CBC mode, every 16-byte segment (4x4 pixel patch) is encypted to 16 bytes of AES digest with AES\_key\_i;
\EndFor
\State re-assemble image blocks $E(Y_i)$ and return $E(Y)$
\end{algorithmic}
\end{algorithm}


\subsection{Complexity Analysis}\label{sec:complexity}
The additional costs of the disguising methods consist of the encoding cost and the possible additional learning cost, i.e., it may take more rounds to converge. We leave the second part to the experimental evaluation and analyze the encoding cost here. 

For an image partitioned into $t$ blocks with each block $l \times m$, the RMT transformation involves $t$ matrix-matrix multiplications and matrix additions. As the numbers $t$, $m$, and $l$ are all small, the cost of RMT per image is low: $O(tlm^2)$. For an image of $l \times m$ with a scale-up constant of $s$, the AES-128 encryption cost is $l \times m \times s$/16 times of AES encryption. 
Our experimental evaluation shows that per image cost is less than 10 ms and can be comfortably done by any PC or mobile phone. 

\subsection{Model Protection via Image Disguising}
Note that the models trained with disguised data work only on disguised data. We show this property also protects models from existing model-targeted attacks. So far, we have seen model-inversion attacks \cite{fredrikson15,zhang20}, membership-inference attacks \cite{shokri16,hu22}, and model-extraction attacks \cite{tramer16,jagielski2020}. 

Model-extraction attacks assume the attacker can freely access the model, e.g., via a cloud-based prediction API. With such a service, the attacker can try various images to collect their outputs and then use the input-output pairs to reconstruct the model. Our threat model assumes the attacker can copy or save the trained model for analysis. Thus, the attacker does not need to perform model-extraction attacks. As the models only work on disguised test images, without the secret disguising key, they are useless to the attacker. 

Membership-inference attack aims to estimate the possibility of a target example belonging to the training data of a model. To perform such an attack, the attacker must first apply the disguising method (with the secret key) to the target example so that the model can be used. This step effectively blocks the attack or at least significantly increases the difficulty. To successfully conduct the MIA attack on the disguised model, the attacker may need to manipulate an authorized user to transform the example and intercept the transformed one, which corresponds to the mentioned Level-2 knowledge. Thus, the disguising mechanism establishes an effective defensive line. 

Model-inversion attack uses a learning procedure, e.g., a GAN method \cite{zhang20}, to progressively adjust randomly generated or seed images from similar domains towards most likely training examples. When applied to the models trained on disguised data, the model-inversion attack recovers only the disguised training data, not the original data. Again, the disguising mechanism builds a defensive line on this attack. We will show how the RMT mechanism works against model-inversion attacks in experiments. 

\section{Attack Analysis}\label{sec:security_analysis}
This section aims to analyze the possible threats to the proposed disguising mechanisms and clarify the applicable settings. With Level-1 adversarial knowledge, Disguised-Nets mechanisms provide strong confidentiality protection, as shown in the discussion of ``brute-force attacks''. In contrast, other related methods are still struggling with visual re-identification by human eyes \cite{li17,fan18} or disguised-image-based reconstruction attacks \cite{carlini21}. We also analyze more sophisticated reconstruction attacks that depend on Level-2 adversarial knowledge. 

\subsection{Level-1 Adversarial Knowledge and Attacks}
Recall that Level-1 knowledge includes knowing the disguised images and possibly the model domain, i.e., the types of images and the background application. It is clear that with only Level-1 knowledge, the brute-force attack on AES schemes is not possible, and thus we focus on the scheme using multidimensional projection. 

\textbf{Visual Re-identification.} The first simple attack is to visually identify images by human attackers. We have shown that simple methods like noise addition, morphing, and shallow-network-based transformation are not resilient to this attack. However, many other attacks may use re-identification as the last step, i.e., reconstruction attacks. It's inefficient for human evaluators to check each image to determine the protection level of an image disguising mechanism. Thus, we propose the \emph{DNN examiner} approach for evaluation purposes: let a DNN trained on the original datasets to perform the visual re-identification task for human evaluators. We will use DNN examiners in experiments. 

\textbf{Brute-Force.} The brute-force attack method for image reconstruction is to enumerate each possible parameter setting of the disguising mechanism and then check the recovered result with re-identification. As AES encryption is already resilient to the brute-force attack, we examine the RMT method only. Let's start with a block-level transformation for any image block $i$ with RMT. With $X_i' = X_iR_i$, the adversary knows only $X_i'$. In the brute force attack, the number of possible $X_i$ is determined by the number of possible $R_i$ matrices. We show that the number of possible $R_i$ (even limited to orthogonal ones) can be exponentially large for given parameters. 
\begin{prop} For values encoded in $h$-bit finite field, there are $O(2^{hm})$ candidate orthogonal matrices $R_{m\times m}$. 
\end{prop}
The proof is based on the theory of orthogonal matrix group \cite{heiberger78}, the detail of which is skipped here. With a typical setting in our experiments, e.g., $h=8$ and $m=28$ for the MNIST dataset, the overall complexity is $O(2^{224})$, which is more than sufficient to protect from computationally-bounded attackers. Combined with the random permutation of blocks, the attack complexity is even higher. Thus, a brute-force attack is generally impractical for the proposed methods.

\textbf{Clustering Attack.}
Carlini et al. \cite{carlini21} utilized a clustering method to attack InstaHide \cite{huang20} disguised images. InstaHide uses the random mix-up method to generate disguised images. Depending on the random weight distribution, some disguised images might be dominated by the same image, which likely forms a cluster of images that can be used to de-mask and de-noise. As InstaHide disguised images are essentially linear combinations of plaintext images, the attack result can be visually re-identified. 

\begin{figure}[h]
\centering
\subfigure[RMT on MNNST]{\includegraphics[scale=.6]{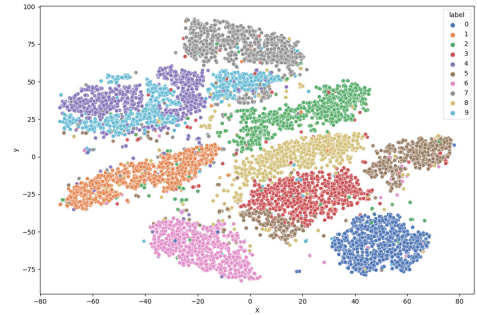}}
\subfigure[RMT on CIFAR10]{\includegraphics[scale=.6]{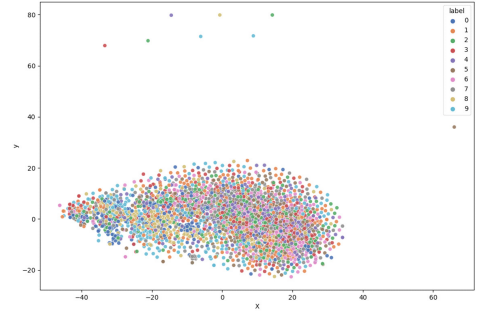}}
\caption{t-SNE visualization of RMT disguised datasets (4x4 blocks). Colors represent different labels. A dense area covered with one color means that the clustering structure matches the label distribution well for the specific subset.}\label{fig:RMT-cluster-vis}
\end{figure}

\begin{figure}[h]
\centering
\subfigure[AES on MNIST]{\includegraphics[scale=.6]{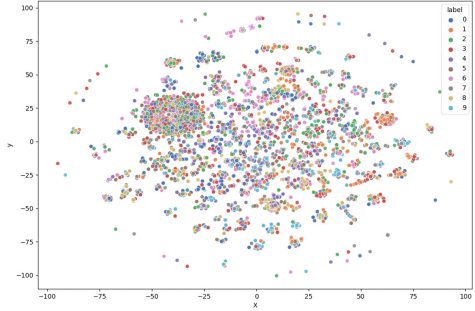}}
\subfigure[AES on CIFAR10]{\includegraphics[scale=.6]{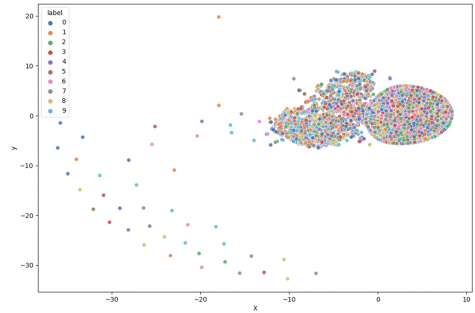}}
\caption{t-SNE visualization of AES disguised datasets (4x4 blocks)}\label{fig:AES-cluster-vis}
\end{figure}

Important questions are whether our disguising methods can generate images with clustering structures and whether such clusters can be used to break our disguising methods. To answer these questions, we visualize the disguised training data with t-SNE \cite{maaten08} to understand the existence of clustering structure in the Euclidean-distance space.   Figure \ref{fig:RMT-cluster-vis} shows that RMT might preserve the clustering structures for some datasets: for simpler datasets like MNIST and FASHION, the clustering structure is well preserved, while others do not. In contrast, AES does not preserve any clustering structure, as shown in Figure \ref{fig:AES-cluster-vis}. While AES not preserving clustering structures to leave less information to attackers, it also affects data utility and leads to lower-quality models, as we will show in experiments. 

Next, can such preserved clustering structures be used for attacks? An attack on InstaHide \cite{carlini21} has used image clusters to de-noise and de-mask, as InstaHide uses the random weights mix-up mechanism. However, unlike InstaHide, clustering structures of RMT-disguised images do not help attackers identify original images. However, it may help attackers infer additional information with other domain knowledge. For example, if the original training samples' distribution (and thus the clustering structure) is also known,  it may allow the attacker to identify the mapping between a specific cluster of original images and a cluster of disguised images. As distances between samples are not preserved, it's still difficult to figure out the sample-to-sample mapping.

\subsection{Level-2 Adversarial Knowledge and Attacks} 
We move one more step further to study the more challenging issue: what if a powerful adversary can obtain additional knowledge: pairs of original images and their transformed ones? This assumption corresponds to the chosen-plaintext attack in cryptographic analysis \cite{katz07}. This study helps us understand when we should not use a proposed disguising method. Below, we focus on the \emph{codebook attack} on the AES-ECB-based disguising mechanism and the regression-based attack on the RMT mechanism.  

\textbf{Codebook Attack.} The assumption is that the adversary is knowledgeable of the encryption procedure described previously but does not have the encryption/decryption key. Since the AES ECB method is deterministic, the basic attack is to build a mapping (i.e., the codebook) between the plaintext unit (e.g., the 16-byte pixel block) and its encrypted counterpart (e.g., the 16-byte AES cipher block). By processing the known image pairs, the adversary constructs a codebook as a dictionary mapping 16-byte pixel blocks to encrypted 16-byte blocks. Since different images, especially those in the same class, might share some 16-byte pixel blocks, some 16-byte encrypted blocks in the targeted images are likely already in the codebook, which will be used to recover the original blocks. For encrypted pixel blocks not present in the codebook, the adversary may use a fixed pattern, e.g., all zero values or most likely values to pad. By repeating this procedure for each 16-byte block, the adversary can recover some parts of the image, which can be further re-identified via human eyes or models at the adversary's hands.

\emph{Possible mitigation methods.} Let the \emph{hit rate} be defined as the probability that an encrypted pixel block can find a match in the codebook. 
This attack can become less effective if we add salt-and-pepper noises to the original images before encoding. This step will reduce the hit rate significantly and make the mapping non-unique: the same 16-byte pixel block can be mapped to different ciphertexts. We will evaluate the \emph{success rate} of this attack in experiments, using the accuracy that the DNN examiner trained with the original image space correctly classifies the reconstructed images.

\textbf{Projection Matrix Estimation Attack.} Note that noise addition can easily defend the RMT method from the codebook attack, which is already a part of the RMT method. However, if the adversary has obtained enough original and transformed image pairs, there is a possibility that the transformation matrix might be estimated with linear regression.     
Specifically, a noise-added block-wise transformation, e.g., $Y_i = R_i(X_i + \Delta_i)$, where $\Delta_i$ is a random noise matrix, re-generated for each image block $X_i$, and drawn uniformly at random from $[0, N]$ where $N$ is the tunable noise level. With enough known pairs of $(X_i, Y_i)$, the regression method can be applied to estimate $R_i$. Generally, the more known pairs, the more precise the estimation can be. However, it's unclear how the noise level affects the effectiveness of estimation and how we can achieve a good balance between data utility and attack resilience. We will examine the regression-based attacks in the experiments.

Note that the recently proposed InstaHide \cite{huang20} method also somewhat matches this definition of image disguising. It also requires learning from the disguised examples $\{(T(X_i), c_i)\}$. However, the learned model $\mathbf{M}_T$ is still applicable to the original test data, i.e., the application phase uses $\mathbf{M}_T(X_{new})$. They also show that the performance of $\mathbf{M}_T(X_{new})$ is very close to $\mathbf{M}(X_{news})$, which implies $\mathbf{M}_T \approx \mathbf{M}$. leads to serious problems, such as the impossibility result and a clustering-based attack, as Carlini et al. \cite{carlini21} show. In contrast, our proposed methods require strictly $\mathbf{M}_T(T(X_{new})$ in the application phase, which eliminates the possible information leakage targeting the models and the clustering of disguised training images.

\section{Experiments}\label{sec:experim}
The experiments have three goals. (1) The proposed DisguisedNets mechanisms involve parameter settings, which may affect data utility. (2) While the proposed methods are resilient to attacks under Level-1 knowledge, we need to understand the intrinsic trade-offs between data utility and the methods enhancing the resilience to Level-2 attacks. (3) As we have discussed, the proposed methods have unique benefits in defending model-based attacks, which we will demonstrate in experiments.

\textbf{Datasets.} We use four prevalent DNN benchmarking datasets: MNIST, FASHION, CIFAR10, and LFW \cite{lfw} for experiments. 
MNIST (handwritten digits) and FASHION (fashion items) are gray-scale $28 \times 28$-pixel images with ten classes. CIFAR10 has 60 thousand $32\times 32$ color images distributed into ten classes. LFW is a labeled face database. It is relatively small, with only a few thousand samples.
We used five folds of random sampling to estimate the standard deviation of modeling results, which are also used for later experiments. 

Table \ref{tab:datasets} summarizes the datasets, the techniques used to train the base models, and their baseline model accuracy on the original image data. All the models are implemented with PyTorch.


\begin{table}[h]
  	\centering
  	\scriptsize
  	\caption{Datasets and Baseline Accuracy. Tr: Training, Te: Testing} 
  	\label{tab:datasets}
  		\begin{tabular}{|c|c|c|c|c|}
  			\hline
  			Datasets & Records& ImageSize & Network& BaselineAccuracy\\
  			\hline
			MNIST & (60K Tr,10K Te) &\{$28 \times 28$\} &AlexNet & $96.7\pm 0.2\%$\\
			FASHION & (60K Tr,10K Te) &\{$28 \times 28$\} &AlexNet & $88.7 \pm 0.3\%$\\
			CIFAR-10 & (50K Tr,10K Te)  & \{$32 \times 32$\} &ResNet-18& $93.4 \pm 0.2\%$\\
			LFW & (1164 Tr, 292 Te)  & \{$60 \times 48$\}  & ResNet-18& $94.3 \pm 2.0\%$\\ 
			\hline
  		\end{tabular}
 \end{table}

\subsection{Parameter Settings for Level-1 Attacks}
Since all the proposed methods are resilient to Level-1 attacks, we focus on the utility preservation aspect in this section.

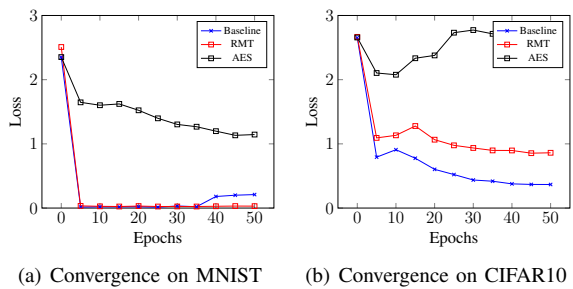
\begin{figure}
\centering
 	\subfigure[Convergence on MNIST]{
		\begin{tikzpicture}[scale=0.45]
		\pgfplotsset{every axis legend/.append style={font=\small},every node near coord/.append style={font=\Large}}
		\begin{axis}
			[ymin=0.00,ymax=3.0,
			xlabel={Epochs},xlabel style = {font = \Large},
			ylabel={Loss}, ylabel style = {font=\Large},
			legend pos=north east,
			y tick label style = {font = \Large},
			x tick label style = {font = \Large}
			]
			\addplot+[mark=x,blue,draw opacity=1]
			table[x=Epochs,y=Origin, col sep=comma]
			{mnist-converge.csv};
			\addlegendentry{Baseline}
						
			\addplot+[mark=square,red,draw opacity=1]
			table[x=Epochs,y=RMT, col sep=comma]
			{mnist-converge.csv};
			\addlegendentry{RMT}
			
			\addplot+[mark=square,black,draw opacity=1]
			table[x=Epochs,y=AES,col sep=comma]
			{mnist-converge.csv};
			\addlegendentry{AES}
		\end{axis}
	\end{tikzpicture}
	}
 	\subfigure[Convergence on CIFAR10]{
		\begin{tikzpicture}[scale=0.45]
		\pgfplotsset{every axis legend/.append style={font=\small},every node near coord/.append style={font=\Large}}
		\begin{axis}
			[ymin=0.00,ymax=3,
			xlabel={Epochs},xlabel style = {font = \Large},
			ylabel={Loss}, ylabel style = {font=\Large},
			legend pos=north east,
			y tick label style = {font = \Large},
			x tick label style = {font = \Large}
			]
			\addplot+[mark=x,blue,draw opacity=1]
			table[x=Epochs,y=Origin, col sep=comma]
			{cifar-converge.csv};
			\addlegendentry{Baseline}
						
			\addplot+[mark=square,red,draw opacity=1]
			table[x=Epochs,y=RMT, col sep=comma]
			{cifar-converge.csv};
			\addlegendentry{RMT}
			
			\addplot+[mark=square,black,draw opacity=1]
			table[x=Epochs,y=AES,col sep=comma]
			{cifar-converge.csv};
			\addlegendentry{AES}
		\end{axis}
	\end{tikzpicture}
	}
 	\caption{Convergence speed on disguised images. Baseline: models trained on original datasets.}
 	\label{fig:convergence}
\end{figure}

\textbf{Costs.} The disguised images are used directly with the existing DNN training algorithms without any modification to the algorithm or data. We have briefly analyzed the per image disguising cost in Section \ref{sec:complexity}, which can be comfortably handled by a mobile phone. Another question is whether the disguised images will extend the training time. Fig \ref{fig:convergence} shows the evaluation of convergence speed on MNIST and CIFAR10 for the three methods: the baseline, RMT, and AES. The baseline refers to the models reported in Table \ref{tab:datasets}. Both RMT and AES run with the basic setting of 4x4 blocks. All of the methods converge with 50 epochs, but AES appears more unstable on CIFAR10. 

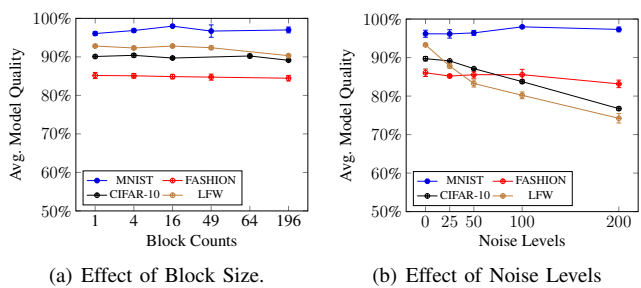
\begin{figure}[h]
\centering
\subfigure[Effect of Block Size.]{
\begin{tikzpicture}[scale=0.45]
 	\begin{axis}
 	[ymin=50,ymax=100.0,
	xlabel={Block Counts},xlabel style = {font = \Large},
 	point meta ={y*100},
 	ylabel={Avg. Model Quality}, ylabel style = {font=\Large},
	symbolic x coords={1,4,16,49,64,196,256},
	yticklabel=\pgfmathprintnumber\tick\%,yticklabel={\pgfmathparse{\tick}\pgfmathprintnumber{\pgfmathresult}\%},
	legend columns=2,
 	legend pos=south west,
	y tick label style = {font = \Large},
	x tick label style = {font = \Large}
	]
	\addplot+[mark=*,error bars/.cd,
 	x dir=both
 	,y dir=both,y explicit]
 	table[x=block_count,y=accuracy, y error = std,col sep=comma, row sep=crcr]
 	{
block_count,accuracy,std\\
196,96.994,0.758307326\\
49,96.698,1.603923315\\
16,97.976,0.347821793\\
4,96.846,0.465327841\\
1,96.054,0.503219634\\
 	};
 	\addlegendentry{MNIST}	
	\addplot+[mark=oplus,color=red,draw opacity=1.0,error bars/.cd,
 	x dir=both
 	,y dir=both,y explicit]
 	table[x=block_count,y=accuracy, y error = std,col sep=comma,row sep=crcr]
 	{
 	block_count,accuracy,std\\
196,84.46,0.73\\
49,84.74,0.79\\
16,84.88,0.57\\
4,85.09,0.61\\
1,85.17,0.79\\
};
 	\addlegendentry{FASHION}	
 	\addplot+[mark=oplus,color=black,draw opacity=1.0,error bars/.cd,
 	x dir=both
 	,y dir=both,y explicit]
 	table[x=block_count,y=accuracy, y error = std,col sep=comma,row sep=crcr]
 	{
block_count,accuracy,std\\
196,89.11666667,0.243\\
64,90.23,0.478\\
16,89.7,0.334\\
4,90.4,0.45\\
1,90.1,0.365\\
 	};
 	\addlegendentry{CIFAR-10}	
	\addplot+[mark=oplus,color=brown,draw opacity=1.0,error bars/.cd,
 	x dir=both
 	,y dir=both,y explicit]
 	table[x=block_count,y=accuracy, y error = std,col sep=comma,row sep=crcr]
 	{
block_count,accuracy,std\\
196,90.32,0.324\\
49,92.372,0.477\\
16,92.811,0.387\\
4,92.313,0.419\\
1,92.813,0.322\\
 	};
 	\addlegendentry{LFW}	
	\end{axis}
 \end{tikzpicture} 
}
\subfigure[Effect of Noise Levels]{
 \begin{tikzpicture}[scale=0.45]
        \begin{axis}
        [ymin=50,ymax=100.0,
        xlabel={Noise Levels},xlabel style = {font = \Large},
        point meta ={y*100},
        ylabel={Avg. Model Quality}, ylabel style = {font=\Large},
        xtick=data,
        yticklabel=\pgfmathprintnumber\tick\%,yticklabel={\pgfmathparse{\tick}\pgfmathprintnumber{\pgfmathresult}\%},
        legend columns=2,
        legend pos=south west,
        y tick label style = {font = \Large},
        x tick label style = {font = \Large}
        ]       
        \addplot+[mark=*,error bars/.cd,
        x dir=both
        ,y dir=both,y explicit]
        table[x=noise_lvl,y=accuracy, y error = std,col sep=comma]
        {./datavp/noise_variation_mnist.csv};
        \addlegendentry{MNIST} 
        \addplot+[mark=oplus,color=red,draw opacity=1.0,error bars/.cd,
        x dir=both
        ,y dir=both,y explicit]
        table[x=noise_lvl,y=accuracy, y error = std,col sep=comma]
        {./datavp/noise_variation_fashion.csv};
        \addlegendentry{FASHION}
           \addplot+[mark=oplus,color=black,draw opacity=1.0,error bars/.cd,
        x dir=both
        ,y dir=both,y explicit]
        table[x=noise_lvl,y=accuracy, y error = std,col sep=comma]
        {./datavp/noise_variation_Cifar.csv};
        \addlegendentry{CIFAR-10}
        \addplot+[mark=oplus,color=brown,draw opacity=1.0,error bars/.cd,
        x dir=both
        ,y dir=both,y explicit]
        table[x=noise_lvl,y=accuracy, y error = std,col sep=comma]
        {./datavp/noise_variation_LFW.csv};
        \addlegendentry{LFW}
      \end{axis}
 \end{tikzpicture}
 }
 	\caption{Effects of block size and varying noise levels on model quality for RMT-disguised images.} 
 	\label{fig:rmt-quality}
\end{figure}

\textbf{RMT Mechanism.} We look at the effects of block size and noise levels  on models trained on images transformed with RMT methods. For easier presentation, we  convert block size into the number of blocks: 1 block on the x-axis means the image is not split into blocks; while 196 blocks means 196 2x2 blocks for 32x32 images (CIFAR10) or padded 28x28 images (MNIST and FASHION), and 196 4x3 blocks for padded 60x48 images (LFW). Thus, a smaller block size results in a larger number of blocks after partitioning, as the image size is fixed. If more than one block is generated in partitioning, we also apply a secret block-wise permutation.
Figure ~\ref{fig:rmt-quality} (a) shows that the model quality is slightly decreased with smaller block sizes (more blocks per image). Overall, the model quality is well preserved, only 2-3\% worse than the baseline. It's also understandable that the simpler images, MNIST and FASHION, are more resilient to noise addition and more sophisticated ones are sensitive to noise as shown in Figure \ref{fig:rmt-quality} (b). 

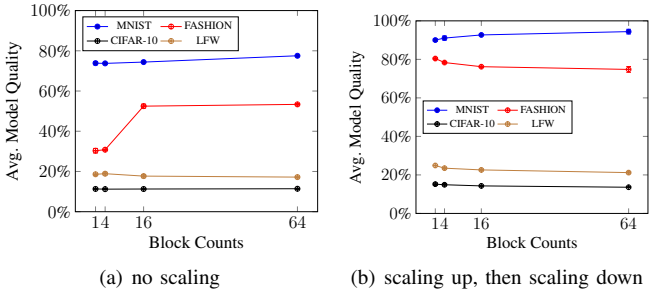
\begin{figure}[h]
\centering
\subfigure[no scaling]{
\begin{tikzpicture}[scale=0.47]
        \pgfplotsset{every axis legend/.append style={font=\small},every node near coord/.append style={font=\Large}}
        \begin{axis}
        [ymin=0.0,ymax=100.0,
        xlabel={Block Counts},xlabel style = {font = \Large},
        point meta ={y*100},
        ylabel={Avg. Model Quality}, ylabel style = {font=\Large},
        xtick=data,
        yticklabel=\pgfmathprintnumber\tick\%,yticklabel={\pgfmathparse{\tick}\pgfmathprintnumber{\pgfmathresult}\%},
        legend columns=2,legend style={
                }, legend pos= north west,
        y tick label style = {font = \Large},
        x tick label style = {font = \Large}
        ]
        \addplot+[mark=*,error bars/.cd,
        x dir=both
        ,y dir=both,y explicit]
        table[x=block_count,y=accuracy, y error = std,col sep=comma,row sep=crcr]
        {
        block_count,accuracy,std\\
        64,77.53,0.25\\
        16,74.39,0.33\\
        4,73.74,0.32\\
        1,73.85,0.8\\
				};
        \addlegendentry{MNIST} 
        \addplot+[mark=oplus,color=red,draw opacity=1.0,error bars/.cd,
        x dir=both
        ,y dir=both,y explicit]
        table[x=block_count,y=accuracy, y error = std,col sep=comma,row sep=crcr]
        {
        block_count,accuracy,std\\
				64,53.36,0.45\\
				16,52.48,1\\
				4,30.83,0.33\\
				1,30.33,1.13\\
        };
        \addlegendentry{FASHION}    
        \addplot+[mark=oplus,color=black,draw opacity=1.0,error bars/.cd,
        x dir=both
        ,y dir=both,y explicit]
        table[x=block_count,y=accuracy, y error = std,col sep=comma,row sep=crcr]
        {
        block_count,accuracy,std\\
64,11.42,0.33\\
16,11.28,0.41\\
4,11.21,0.31\\
1,11.32,0.41\\
        };
        \addlegendentry{CIFAR-10}       
         \addplot+[mark=oplus,color=brown,draw opacity=1.0,error bars/.cd,
        x dir=both
        ,y dir=both,y explicit]
        table[x=block_count,y=accuracy, y error = std,col sep=comma,row sep=crcr]
        {
        block_count,accuracy,std\\
64,17.2,0.57\\
16,17.7,0.78\\
4,18.9,0.66\\
1,18.6,0.59\\
        };
        \addlegendentry{LFW}      
           
        \end{axis}
 \end{tikzpicture}
 }
 \subfigure[scaling up, then scaling down]{
\begin{tikzpicture}[scale=0.45]
        \pgfplotsset{every axis legend/.append style={font=\small},every node near coord/.append style={font=\Large}}
        \begin{axis}
        [ymin=0.00,ymax=100.0,
        xlabel={Block Counts},xlabel style = {font = \Large},
        point meta ={y*100},
        ylabel={Avg. Model Quality}, ylabel style = {font=\Large},
        xtick=data,
        yticklabel=\pgfmathprintnumber\tick\%,yticklabel={\pgfmathparse{\tick}\pgfmathprintnumber{\pgfmathresult}\%},
        legend columns=2,legend style={at={(0.03,0.5)},anchor=west,},
        y tick label style = {font = \Large},
        x tick label style = {font = \Large}
        ]       
        \addplot+[mark=*,error bars/.cd,
        x dir=both
        ,y dir=both,y explicit]
        table[x=block_count,y=accuracy, y error = std, col sep=comma, row sep=crcr] {
        	block_count,accuracy,std\\
					1,90.03,0.74\\
					4,91.06,1.07\\
					16,92.69,0.31\\
					64,94.41,1.2\\
        };
        \addlegendentry{MNIST} 
                \addplot+[mark=oplus,color=red,draw opacity=1.0,error bars/.cd,
        x dir=both
        ,y dir=both,y explicit]
        table[x=block_count,y=accuracy, y error = std,col sep=comma,row sep=crcr]
        {
					block_count,accuracy,std\\
					64,74.78,1.5\\
					16,76.2,0.42\\
					4,78.35,0.52\\
					1,80.43,0.46\\
        };
        \addlegendentry{FASHION} 
         \addplot+[mark=oplus,color=black,draw opacity=1.0,error bars/.cd,
        x dir=both
        ,y dir=both,y explicit]
        table[x=block_count,y=accuracy, y error = std,col sep=comma,row sep=crcr]
        {
        block_count,accuracy,std\\
				64,13.6,0.23\\
				16,14.3,0.36\\
				4,14.9,0.62\\
				1,15.2,0.68\\
        };
        \addlegendentry{CIFAR-10}   
                \addplot+[mark=oplus,color=brown,draw opacity=1.0,error bars/.cd,
        x dir=both
        ,y dir=both,y explicit]
        table[x=block_count,y=accuracy, y error = std,col sep=comma,row sep=crcr]
        {
					block_count,accuracy,std\\
					64,21.2,0.62\\
					16,22.6,0.54\\
					4,23.5,0.57\\
					1,24.9,0.24\\
        };
        \addlegendentry{LFW}   
      
        \end{axis}
 \end{tikzpicture}
 }
        \caption{Effect of block size on model quality for AES-ECB-disguised images}
        \label{fig:blocksize_aes}
\end{figure}

\textbf{AES Mechanism.} We have done experiments to understand the effect of block-size setting for the AES ECB based block protection. We use ``pixel blocks'' for partitioning and permutation, and ``units'' for AES encryption units. A pixel block typically contains more than one unit. Recall that AES uses 16 bytes as the encryption unit if 128-bit encryption is used. Our partitioning schemes follow this restriction of unit size to make sure that each block has integer times of 16 bytes. Figure \ref{fig:blocksize_aes} shows different block size settings from 1 block (e.g., 32x32 per block for 32x32 images) to 64 blocks (e.g., 4x4 pixels per block for 32x32 images). 

We tested two schemes: no scaling vs scaling. The no-scaling scheme uses the block size $\geq$ 16 bytes, while scaling can use even smaller block sizes. Specifically, when we use a block size < 16, e.g., 2x2 blocks, the scaling up factors are determined for the $x$ and $y$ axes, corresponding, e.g., the scaling factor for x-axis is 2 and also 2 for y-axis for 2x2 blocks, so that we can partition the scaled image with 4x4 blocks. Figure \ref{fig:blocksize_aes} (a) shows that the model quality can be affected by the no-scaling scheme. For some datasets, e.g., CIFAR10 and LFW, the model quality is too low to be used. Figure \ref{fig:blocksize_aes} (b)  shows that the model quality is boosted to the level comparable to the RMT's results for MNIST and FASHION, while the other two still stay at unusable levels. The possible reason is that the colored (multi-channel) images contain more noisy image blocks, which changes significantly after the AES transformation. In summary, different from the RMT scheme, the AES scheme may only work for some datasets.

\subsection{Resilience on Level-2 Attacks: AES Scheme}

With the known additional knowledge, i.e., pairs of original and disguised images, the disguising mechanisms might be under the reconstruction attack, and attackers can visually check the reconstructed images to re-identify the features of original images. 

To effectively evaluate the re-identification step, we use a DNN trained on the original image data to simulate the attacker in the visual re-identification process. The intuition is that if any features in the disguised (or reconstructed) images can be detected visually by the adversary, it can be used to link the disguised (or reconstructed) images to the original images. Such linking is often probabilistic, and we can use the linking \emph{success rate} (i.e., the accuracy of prediction) to gauge the threat level of attacks. As DNNs perform comparably well as human experts do in the image-based classification tasks ~\cite{krizhevsky17}, we believe such ``DNN examiners'' can satisfactorily simulate the attacker.  

We train DNN examiners with the original training data using the same DNN architectures detailed in Table \ref{tab:datasets}. We then apply the DNN examiners to see whether the reconstructed images can be correctly classified to their original labels.  To minimize the impact of DNN architecture and different baseline accuracy, we define the \emph{attack success rate} as  
\[\frac{\text{\footnotesize accuracy of DNN examiner on disguised/reconstructed images}}{ \text{\footnotesize accuracy of DNN examiner on original images}} \times 100.\]

\subsubsection{Resilience to Codebook Attack for the AES ECB method}
Assume the attacker knows $m$ pairs of original images and their ECB encrypted ones, and also other information such as their pixel-block sizes. The codebook attack uses the known pairs to construct a mapping between the known plaintext 16-byte pixels (or a reduced number of pixels if the scaling up/down method is used to preserve more utility) and the corresponding encrypted 16-byte pixels. The attacker might be able to use the codebook to partially recover the original pixel blocks of a disguised image (with random pixel patches for unrecognized blocks). We use the DNN examiner to examine the quality of reconstructed images. 

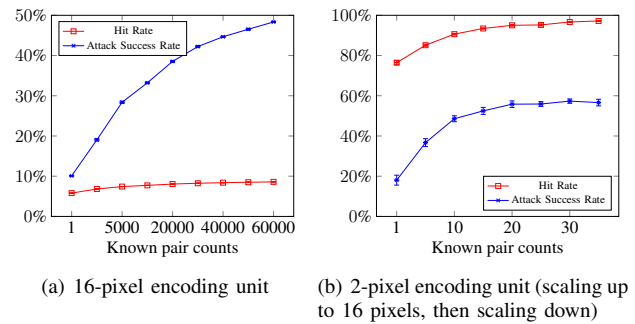
\begin{figure}[h]
	\centering
	\subfigure[16-pixel encoding unit]{
		\begin{tikzpicture}[scale=0.47]
		\pgfplotsset{every axis legend/.append style={font=\small},every node near coord/.append style={font=\Large}}
		\begin{axis}
			[ymin=0.0,ymax=50.0,
			xlabel={Known pair counts},xlabel style = {font = \Large},
			point meta ={y*100},
			ylabel={}, ylabel style = {font=\Large},
			symbolic x coords={1, 1000, 5000, 10000, 20000, 30000, 40000, 50000, 60000},
			yticklabel=\pgfmathprintnumber\tick\%,yticklabel={\pgfmathparse{\tick}\pgfmathprintnumber{\pgfmathresult}\%},
			legend pos=north west,
			y tick label style = {font = \Large},
			x tick label style = {font = \Large}
			]
			\addplot+[mark=square,red,draw opacity=1,error bars/.cd,
			x dir=both,
			y dir=both,y explicit]
			table[x=known_pair,y=hitrate,y error=std, col sep=comma,row sep=crcr]
			{
			known_pair,hitrate,std\\
1,5.80319922,0.00347734\\
1000,6.82191016,0.01065582\\
5000,7.41815625,0.00314082\\
10000,7.72047656,0.00253262\\
20000,8.0475,0.00363871\\
30000,8.24737109,0.00087995\\
40000,8.389625,0.00118944\\
50000,8.50214063,0.00051241\\
60000,8.59485937,8.73E-06\\
			};
			\addlegendentry{Hit Rate}
			\addplot+[mark=x,blue,draw opacity=1,error bars/.cd,
			x dir=both,
			y dir=both,y explicit]
			table[x=known_pair,y=success,y error=std, col sep=comma, row sep=crcr]
			{
			known_pair,success,std\\
1,10.08893485,0.02015883\\
1000,19.04033092,0.305583844\\
5000,28.38055843,0.293535213\\
10000,33.22026887,0.201269752\\
20000,38.5336091,0.233607246\\
30000,42.22957601,0.258676278\\
40000,44.66597725,0.164585329\\
50000,46.53154085,0.230936835\\
60000,48.34539814,0.12845662\\
			};
			\addlegendentry{Attack Success Rate}			
		\end{axis}
	\end{tikzpicture}
	}
	\subfigure[2-pixel encoding unit (scaling up to 16 pixels, then scaling down)]{
		\begin{tikzpicture}[scale=0.47]
		\pgfplotsset{every axis legend/.append style={font=\small},every node near coord/.append style={font=\Large}}
		\begin{axis}
			[ymin=0.0,ymax=100.0,
			xlabel={Known pair counts},xlabel style = {font = \Large},
			point meta ={y*100},
			ylabel={}, ylabel style = {font=\Large},
			symbolic x coords={1, 5, 10, 15, 20, 25, 30, 35},
			yticklabel=\pgfmathprintnumber\tick\%,yticklabel={\pgfmathparse{\tick}\pgfmathprintnumber{\pgfmathresult}\%},
			legend pos=south east,
			y tick label style = {font = \Large},
			x tick label style = {font = \Large}
			]		
			\addplot+[mark=square,red,draw opacity=1,error bars/.cd,
			x dir=both,
			y dir=both,y explicit]
			table[x=known_pair,y=hitrate,y error=std, col sep=comma, row sep=crcr]
			{
			known_pair,hitrate,std\\
1,76.4480352,0.98748169\\
5,85.1324219,0.5004759\\
10,90.6220664,0.38617351\\
15,93.4500234,0.33798848\\
20,95.0233906,0.40677503\\
25,95.2123203,0.50214757\\
30,96.6785586,0.05247563\\
35,97.1809727,0.15382241\\
};
			\addlegendentry{Hit Rate}
			\addplot+[mark=x,blue,draw opacity=1,error bars/.cd,
			x dir=both,
			y dir=both,y explicit]
			table[x=known_pair,y=success,y error=std, col sep=comma, row sep=crcr]
			{
			known_pair,success,std\\
1,18.026,2.4624033\\
5,36.69,1.96795579\\
10,48.576,1.46442139\\
15,52.468,1.7560951\\
20,55.814,1.63691173\\
25,55.89,1.1840397\\
30,57.334,1.08264953\\
35,56.618,1.64574299\\
			};
			\addlegendentry{Attack Success Rate}
		\end{axis}
	\end{tikzpicture}
}
	\caption{Codebook attack on MNIST dataset with varying number of known pairs.}
	\label{fig:codebook_attack_mnist}
\end{figure}

As MNIST and Fashion perform reasonably well with the AES scheme (Figure \ref{fig:blocksize_aes} compared to the other two, we pick only the MNIST data for clear presentation -- the Fashion data has a similar pattern. Figure \ref{fig:codebook_attack_mnist} compares the attack results on 16-pixel encryption units (subfigure (a)) and 2-pixel encryption units with scaling (subfigure (b)). The attacker's known pairs are selected randomly from the training data, while the targeted images are selected from the testing data. 16-pixel encryption unit gives a one-to-one mapping between the original pixel units and the encrypted ones. We observed hit rates are quite low (lower than 10\%), but success rates are increasing steadily due to the increased codebook size. Overall, attackers will need a large number of pairs to achieve a good success rate. 2-pixel encryption unit may create a one-to-many mapping between original pixel units and the encrypted ones, due to the scale up/down processes. We used the Python library function for image scaling. With the scaling process, we observed that hit rates initially increase to around 10\% and then drop to 2-3\%. However, the success rate quickly reaches the plateau -- around 50\% with only 20 image pairs. Therefore, no-scaling method is more resilient to attacks -- both the hit rates and success rates grow slowly and knowing the whole training data does not help improve the success rates much. In contrast, the scaling method can help gain better model quality. However, it might be vulnerable to Level-2 attacks.  There seems an abrupt trade-off the user may have to make. 

Aiming at achieving a better balance of utility and attack resilience for the setting of the 2-pixel encryption unit, we found that it's possible to defend from the codebook attack by adding ``salt-and-pepper'' noises to the original images. The AES encrypted pixel block changes dramatically when any of the original pixel changes, which helps reduce the attack success rate. Figure \ref{fig:code_book_noise_mnist} shows by adding a small amount of noise, e.g., 2-3\%, the attack success rate drops by 10\%, while the model quality is not significant damaged. Certainly, the level of noise should be carefully chosen to avoid destroying the data utility: an increase of noise intensity to 4\% will dramatically degrade the model quality as Figure \ref{fig:code_book_noise_mnist} (b) shows. 

\begin{figure}[h]
	\centering
	\subfigure[16-pixel encoding unit]{
	    \begin{tikzpicture}[scale=0.47]
    	\pgfplotsset{every axis legend/.append style={font=\small},every node near coord/.append style={font=\Large}}
    	\begin{axis}
    		[ymin=0.0,ymax=100,
    		xlabel={Noise Level},xlabel style = {font = \Large},
    		point meta ={y*100},
    		ylabel={}, ylabel style = {font=\Large},
    		symbolic x coords={0.5, 1, 2, 3, 4},
    		yticklabel=\pgfmathprintnumber\tick\%,yticklabel={\pgfmathparse{\tick}\pgfmathprintnumber{\pgfmathresult}\%},
    		legend pos=north west,
    		y tick label style = {font = \Large},
    		x tick label style = {font = \Large}
    		]    		
    		\addplot+[mark=x,blue,draw opacity=1,error bars/.cd,
    		x dir=both,
    		y dir=both,y explicit]
    		table[x=noise,y=success,y error=s_std, col sep=comma, row sep=crcr]
    		{
    		noise,hitrate,h_std,success,s_std,model_success,m_std\\
0.5,8.29172921,0.01905736,44.475698,0.64217486,68.808,2.61592622\\
1,8.03754321,0.00916203,41.3050672,0.05442688,71.046,0.56109714\\
2,7.5220054,0.01213925,35.8965874,0.57574922,62.928,1.04702913\\
3,6.95932748,0.03561122,30.2792141,0.16658742,62.142,0.70990844\\
4,6.48824849,0.05808586,24.8810755,0.69054544,56.808,1.134447\\
    		};
    		\addlegendentry{Attack Success Rate}
    		
    		\addplot+[mark=oplus,black,draw opacity=1,error bars/.cd,
    				x dir=both,
    				y dir=both,y explicit]
    				table[x=noise,y=model_success,y error=m_std, col sep=comma, row sep=crcr]
    				{
    				noise,hitrate,h_std,success,s_std,model_success,m_std\\
0.5,8.29172921,0.01905736,44.475698,0.64217486,68.808,2.61592622\\
1,8.03754321,0.00916203,41.3050672,0.05442688,71.046,0.56109714\\
2,7.5220054,0.01213925,35.8965874,0.57574922,62.928,1.04702913\\
3,6.95932748,0.03561122,30.2792141,0.16658742,62.142,0.70990844\\
4,6.48824849,0.05808586,24.8810755,0.69054544,56.808,1.134447\\
};
    				\addlegendentry{Model Quality}
    	\end{axis}
    \end{tikzpicture}
    }
    \subfigure[2-pixel encoding unit (scaling up and then down)]{
	\begin{tikzpicture}[scale=0.47]
		\pgfplotsset{every axis legend/.append style={font=\small},every node near coord/.append style={font=\Large}}
			\begin{axis}
		[ymin=0.0,ymax=100.0,
		xlabel={Noise Level},xlabel style = {font = \Large},
		point meta ={y*100},
		ylabel={}, ylabel style = {font=\Large},
		symbolic x coords={0.5, 1, 2, 3, 4},
		yticklabel=\pgfmathprintnumber\tick\%,yticklabel={\pgfmathparse{\tick}\pgfmathprintnumber{\pgfmathresult}\%},
		legend pos=south west,
		y tick label style = {font = \Large},
		x tick label style = {font = \Large}
		]		
		\addplot+[mark=x,blue,draw opacity=1,error bars/.cd,
		x dir=both,
		y dir=both,y explicit]
		table[x=noise,y=success,y error=std, col sep=comma, row sep=crcr]
		{
		noise,success,std\\
0.5,53.5346432,0.77343374\\
1,42.0144778,0.63682359\\
2,37.377456,0.86695922\\
3,30.7569804,1.78747443\\
4,18.0372285,1.3494336\\
		};
		\addlegendentry{Attack Success Rate}		
		\addplot+[mark=oplus,black,draw opacity=1,error bars/.cd,
		x dir=both,
		y dir=both,y explicit]
		table[x=noise,y=success,y error=std, col sep=comma, row sep=crcr]
		{
		noise,success,std\\
0.5,94.308,1.08024071\\
1,92.532,0.37811374\\
2,91.638,1.10397917\\
3,90.136,0.9228651\\
4,21.158,22.6589523\\
		};
		\addlegendentry{Model Quality}		
		\end{axis}
	\end{tikzpicture}
	}
	\caption{Protecting AES-based disguising with noise addition (MNIST data)}
	\label{fig:code_book_noise_mnist}
\end{figure}
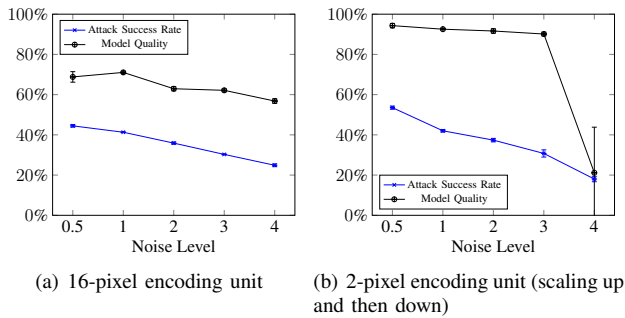

 \begin{figure}[h]
	\centering
	\subfigure[direct re-identification attack is not effective]{
\begin{tikzpicture}[scale=0.45]
 	\pgfplotsset{every axis legend/.append style={font=\small},every node near coord/.append style={font=\Large}}
 	\begin{axis}
 	[ymin=0.0,ymax=40.0,
	xlabel={Block Counts},xlabel style = {font = \Large},
 	point meta ={y*100},
 	ylabel={Attack Success Rate}, ylabel style = {font=\Large},
	symbolic x coords={1,4,16,49,64,196,256},
	yticklabel=\pgfmathprintnumber\tick\%,yticklabel={\pgfmathparse{\tick}\pgfmathprintnumber{\pgfmathresult}\%},
 	legend pos=north east,
	y tick label style = {font = \Large},
	x tick label style = {font = \Large}
	]
	\addplot+[mark=*,blue,error bars/.cd,
 	x dir=both,y dir=both,y explicit]
 	table[x=block_count,y=accuracy,y error=std,col sep=comma, row sep=crcr]
 	{
 	block_count,accuracy,std\\
196,14.24,2.4\\
49,15.78,1.3\\
16,16.23,1.4\\
4,16.48,2.1\\
1,17.32,2.2\\
 	};
 	\addlegendentry{MNIST}
	\addplot+[mark=oplus,color=black,draw opacity=1.0,error bars/.cd,
 	x dir=both,y dir=both,y explicit]
 	table[x=block_count,y=accuracy,y error=std,col sep=comma, row sep=crcr]
 	{
 block_count,accuracy,std\\
196,13.41,2.33\\
49,12.78,2.26\\
16,13.54,1.89\\
4,14.21,2.76\\
1,13.27,2.07\\
};
 	\addlegendentry{FASHION}	

			\addplot+[mark=square,black,draw opacity=1,error bars/.cd,
			x dir=both,
			y dir=both,y explicit]
			table[x=block_count,y=accuracy,y error=std, col sep=comma, row sep=crcr]
			{%
			block_count,accuracy,std\\
256,13.21,1.17\\
64,13.28,0.94\\
16,13.11,1.28\\
4,13.18,1.37\\
1,14.22,1.76\\
			};
			\addlegendentry{CIFAR-10}
			
			\addplot+[mark=square,brown,draw opacity=1,error bars/.cd,
			x dir=both,
			y dir=both,y explicit]
			table[x=block_count,y=accuracy,y error=std, col sep=comma, row sep=crcr]
			{%
			block_count,accuracy,std\\
196,13.41,2.33\\
49,12.78,2.26\\
16,13.54,1.89\\
4,14.21,2.76\\
1,13.27,2.07\\
			};
			\addlegendentry{LFW}

 	\end{axis}
 	\end{tikzpicture}
 	}
 	\subfigure[Regression attacks on the noise-added RMT disguising method can be  effective with enough known pairs. Images with block-size ${7 \times 7}$  and noise level $u=100$]{
		\begin{tikzpicture}[scale=0.45]
		\pgfplotsset{every axis legend/.append style={font=\small},every node near coord/.append style={font=\Large}}
		\begin{axis}
			[ymin=0.00,ymax=120.0,
			xlabel={Known pair counts},xlabel style = {font = \Large},
			point meta ={y*100},
			ylabel={Avg. Attack Success Rate}, ylabel style = {font=\Large},
			symbolic x coords={1, 5, 10, 15, 20, 25, 30, 35},
			yticklabel=\pgfmathprintnumber\tick\%,yticklabel={\pgfmathparse{\tick}\pgfmathprintnumber{\pgfmathresult}\%},
			legend pos=south east,
			y tick label style = {font = \Large},
			x tick label style = {font = \Large}
			]
			\addplot+[mark=x,blue,draw opacity=1,error bars/.cd,
			x dir=both,
			y dir=both,y explicit]
			table[x=known_pair,y=mnist,y error=m_std, col sep=comma]
			{rmt_attack_known_pair_success.csv};
			\addlegendentry{MNIST}
						
			\addplot+[mark=square,red,draw opacity=1,error bars/.cd,
			x dir=both,
			y dir=both,y explicit]
			table[x=known_pair,y=fashion,y error=f_std, col sep=comma]
			{rmt_attack_known_pair_success.csv};
			\addlegendentry{Fashion}
			
			\addplot+[mark=square,black,draw opacity=1,error bars/.cd,
			x dir=both,
			y dir=both,y explicit]
			table[x=known_pair,y=Cifar,y error=c_std, col sep=comma]
			{rmt_attack_known_pair_success.csv};
			\addlegendentry{CIFAR-10}
			
			\addplot+[mark=square,brown,draw opacity=1,error bars/.cd,
			x dir=both,
			y dir=both,y explicit]
			table[x=known_pair,y=LFW,y error=l_std, col sep=comma]
			{rmt_attack_known_pair_success.csv};
			\addlegendentry{LFW}

		\end{axis}
	\end{tikzpicture}
	}
 	\caption{Attacks on RMT-disguised images.}
 	\label{fig:rmt-re-identification}
\end{figure}
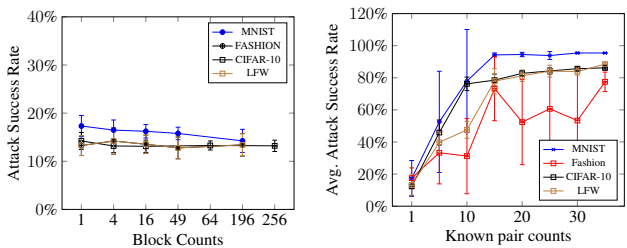

\subsection{Resilience on Level-2 Attacks: RMT Scheme}
We study how known pairs can be effectively used to attack the RMT method. Again, we assume a stronger attack scenario: the attacker already knows the pixel-block size and the permutation pattern. By known only one pair of images, RMT without noise addition can be easily broken -- the block-wise transformation parameters $\{R_i, i=1..m\}$ can be straightforwardly recovered. Different from the ``salt-and-pepper'' noise for selected pixels in the enhanced AES scheme, we generate a noise value for each pixel and add it to the original pixel value before applying the projection, i.e., $Y_i = (X_i + \Delta_i)R_i$, where the noise $\Delta_i$ is drawn from the uniform distribution $U(0, u)$. With noise addition, the known attack method is to use linear regression to estimate the parameters $\{R_i\}$, the accuracy of which is affected by the noise intensity (i.e., the variance of noise) and the number of available pairs. 

Figure \ref{fig:rmt-re-identification} (a) shows that direct re-identification (with Level-1 attack) is generally not effective at all. However, Figure \ref{fig:rmt-re-identification} (b) shows that the regression attack is surprisingly effective on all datasets. With a small number of known image pairs, the attack can achieve surprisingly high success rates. Thus, it's not safe to use the RMT scheme when Level-2 attack knowledge is possibly available.


\pgfplotstableread{
0 0.677 0.008 0.093 0.005 0.123 0.007 0.162 0.012
1 0.543 0.011 0.08 0.002 0.083 0.004 0.091 0.004
2 0.228 0.016 0.021 0.008 0.016 0.004 0.026 0.009
3 0.196 0.023 0.031 0.005 0.042 0.011 0.047 0.008
}\acc
\begin{figure}[h]
\centering
\resizebox{0.6\columnwidth}{!}{
\begin{tikzpicture}[font=\small]
\begin{axis}[ybar,
        width=0.6\textwidth,
        bar width=10pt,
        ymin=0,
        ymax=0.7,        
        ylabel={Attack Success Rate},
        xtick=data,
        label style={font=\small},
        tick label style={font=\small}  ,
        xticklabels = {MNIST,FASHION,LFW,CIFAR-10},
        major x tick style = {opacity=0},
        minor x tick num = 1,
        minor tick length=2ex,
        legend pos=north east
        ]

\addplot[draw=black,fill=black!0,error bars/.cd,y dir=both,y explicit] 
    table[x index=0,y index=1,y error plus index=2,y error minus index=2]\acc; 
    \addplot[draw=black,fill=black!20,error bars/.cd,y dir=both,y explicit] 
     table[x index=0,y index=3,y error plus index=4,y error minus index=4]\acc; 
    \legend{Original,RMT}
\end{axis}
\end{tikzpicture}}
\caption{RMT protects models from model-inversion attacks. Original: the MI attack applied to the original non-protected model to recover images. RMT: the MI attack applied to the model trained on RMT-disguised training data.} \label{fig:mi-attack}
\end{figure}
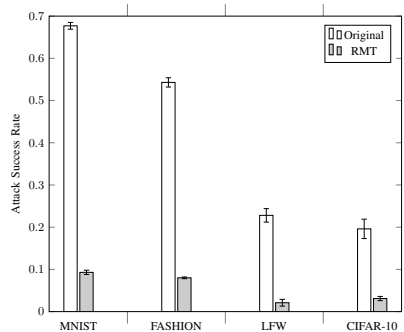 

\subsection{Use Image Disguising to Protect Models}
Exposing models may have high risks, as shown in model-inversion attacks, membership-inference attacks, and model-extraction attacks. This experiment shows that image-disguising methods can work effectively against such model-targeted attacks. We take model-inversion (MI) attacks, for example, which try to recover training data from the exposed model. 

The experiment exposes the models trained on RMT disguised images for a recent model-inversion (MI) attack \cite{zhang20} that has shown good performance in recovering training data. Specifically, we used a 4x4 block without noise addition for RMT to generate disguised data and models. We then apply the MI attack to generate 2000 images for each dataset (200 for each class). To compare the performance of the MI attack, we use the DNN examiners trained on the original datasets to recognize the recovered images. Figure \ref{fig:mi-attack} shows that models trained on RMT-disguised data are very resilient to the MI attack. Indeed, the MI attack recovers the RMT-disguised training data, which are different from the original images and thus still unrecognizable. The results are consistently worse than the DNN examiners applied to the RMT-disguised training data directly (Figure \ref{fig:rmt-re-identification} a).

\subsection{Discussion}
Based on the experimental results, we have the following observations. 
\begin{itemize}
	\item With Level-1 adversarial knowledge, the RMT mechanism preserves good data utility for most datasets. In contrast, the AES scheme only keeps data utility for some datasets. Table \ref{tab:level-1} summarizes the best result under the Level-1 adversarial knowledge assumption. 
\begin{table}[t]
   	\centering
   	\caption{Best Result Under Level-1 Assumption} \label{tab:level-1}
\begin{tabular}{|c|c|c|c|}
 		\hline
 		Datasets & No Disguise &RMT Disguise & AES Disguise\\
   			\hline
 			MNIST &96.7 $\pm$0.2\%&96.6 $\pm$ 0.4 \% &91.6 $\pm$ 1.1\% \\
FASHION& 88.7 $\pm $0.3\% &85.1 $\pm$ 0.6 \%& 68.9$\pm$1.4\%\\
 			CIFAR-10 & 93.4 $\pm$0.2\%& 89.3\%$\pm$0.1\% &11.3 $\pm$ 1.7 \%  \\
 			LFW &94.3 $\pm$2.0&92.6$\pm$2.3\%&17.2$\pm$0.4\% \\
 			\hline
   		\end{tabular}
   \end{table}

\item With Level-2 adversaries, the RMT mechanism should not be used as the attack success rate will be high. The AES scheme with a small encryption unit and small (e.g., 2\%) noise addition is resilient to the codebook attack and still preserves model quality for some datasets. Table \ref{tab:level-2} summarizes the best results for AES. As the AES scheme does not work on CIFAR10 and LFW, so far, we haven't discovered satisfactory utility-preserving disguising methods against Level-2 adversaries.

\begin{table}[t]
   	\centering
   	\caption{AES Best Result Under Level-2 Assumption: encryption unit 2x1 (with scaling), noise level 2\%.} \label{tab:level-2}
\begin{tabular}{|c|c|c|c|}
 		\hline
 		Datasets & No Disguise & Model Accuracy & Attack Success Rate\\
   			\hline
 			MNIST &96.7 $\pm$0.2\%&90.14 $\pm$1.1\% &30.76$\pm$0.87\% \\
FASHION& 88.7 $\pm $0.3\% &73.08$\pm$ 0.86\%&23.51$\pm$ 0.27\% \\
 			\hline
   		\end{tabular}
   \end{table}

\item Finally, if only Level-1 adversaries are expected, RMT can also be used to effectively protect from model-targeted attacks, as the models trained on RMT disguised data can only be applied to disguised data. 
\end{itemize}

\section{Conclusion}\label{sec:conclusion}
Outsourcing large image datasets to the cloud for deep learning has been an economical and popular option, but it also raises concerns about data and model confidentiality. The existing solutions are either too expensive to be practical, vulnerable to different model-based adversarial attacks, or ineffective in protecting the image content. By focusing on the training image reconstruction and re-identification attacks,  we propose image disguising mechanisms that efficiently thwart the attacks and preserve model quality. The combination of random image-block permutation and block-wise AES encryption or multidimensional transformation (RMT) does not require any changes to the existing DNN modeling architectures. Experimental results show that the RMT method can preserve the model quality and provide sufficient attack resilience under Level-1 adversarial knowledge -- adversaries knowing only the disguised images and the domain information. The AES method improves the attack resilience against Level-2 adversaries who manage to obtain pairs of original images and disguised ones. However, the AEs method may seriously damage some datasets' utility. We also show that the disguising methods can protect the trained models from model-targeted attacks.  
The future work will be focused on new image disguising mechanisms that can more efficiently preserve utility with stronger security guarantees. We will also extend the related research to non-image data.

\bibliographystyle{abbrv}
\bibliography{sgx_sc,paper,paper2}

\begin{thebibliography}{10}

\bibitem{abadi16}
M.~Abadi, A.~Chu, I.~Goodfellow, H.~B. McMahan, I.~Mironov, K.~Talwar, and
  L.~Zhang.
\newblock Deep learning with differential privacy.
\newblock In {\em Proceedings of the 2016 ACM SIGSAC Conference on Computer and
  Communications Security}, CCS '16, pages 308--318, New York, NY, USA, 2016.
  ACM.

\bibitem{carlini21}
N.~Carlini, S.~Deng, S.~Garg, S.~Jha, S.~Mahloujifar, M.~Mahmoody, S.~Song,
  A.~Thakurta, and F.~Tram{\`e}r.
\newblock Is private learning possible with instance encoding?
\newblock In {\em IEEE Symposium on Security and Privacy (S\&P)}, 2021.

\bibitem{chakraborty18}
A.~Chakraborty, M.~Alam, V.~Dey, A.~Chattopadhyay, and D.~Mukhopadhyay.
\newblock Adversarial attacks and defences: {A} survey.
\newblock {\em CoRR}, abs/1810.00069, 2018.

\bibitem{chen10}
A.~Chen.
\newblock Gcreep: Google engineer stalked teens, spied on chats.
\newblock {\em Gawker, http://gawker.com/5637234/}, 2010.

\bibitem{duncan12}
A.~J. Duncan, S.~Creese, and M.~Goldsmith.
\newblock Insider attacks in cloud computing.
\newblock In {\em 2012 IEEE 11th International Conference on Trust, Security
  and Privacy in Computing and Communications}, 2012.

\bibitem{dwork06}
C.~Dwork.
\newblock Differential privacy.
\newblock In {\em International Colloquium on Automata, Languages
  andProgramming}, pages 1--12. Springer, 2006.

\bibitem{fan18}
L.~Fan.
\newblock Image pixelization with differential privacy.
\newblock In {\em Data and Applications Security and Privacy {XXXII} - 32nd
  Annual {IFIP} {WG} 11.3 Conference, DBSec 2018, Bergamo, Italy, July 16-18,
  2018, Proceedings}, pages 148--162, 2018.

\bibitem{fei21}
S.~Fei, Z.~Yan, W.~Ding, and H.~Xie.
\newblock Security vulnerabilities of sgx and countermeasures: A survey.
\newblock {\em ACM Comput. Surv.}, 54(6), 2021.

\bibitem{fredrikson15}
M.~Fredrikson, S.~Jha, and T.~Ristenpart.
\newblock Model inversion attacks that exploit confidence information and basic
  countermeasures.
\newblock In {\em ACM Conference on Computer and Communications Security},
  2015.

\bibitem{fredrikson14}
M.~Fredrikson, E.~Lantz, S.~Jha, S.~Lin, D.~Page, and T.~Ristenpart.
\newblock Privacy in pharmacogenetics: An end-to-end case study of personalized
  warfarin dosing.
\newblock In {\em 23rd USENIX Security Symposium USENIX Security}, pages
  17--32, San Diego, CA, 2014. USENIX Association.

\bibitem{gallier00}
J.~Gallier.
\newblock {\em Geometric Methods and Applications for Computer Science and
  Engineering}.
\newblock Springer-Verlag, New York, 2000.

\bibitem{gilad16}
R.~Gilad-Bachrach, N.~Dowlin, K.~Laine, K.~Lauter, M.~Naehrig, and J.~Wernsing.
\newblock Cryptonets: Applying neural networks to encrypted data with high
  throughput and accuracy.
\newblock In M.~F. Balcan and K.~Q. Weinberger, editors, {\em Proceedings of
  The 33rd International Conference on Machine Learning}, volume~48 of {\em
  Proceedings of Machine Learning Research}, pages 201--210, 2016.

\bibitem{heiberger78}
R.~M. Heiberger.
\newblock Generation of random orthogonal matrix.
\newblock {\em Journal of the Royal Statistical Society}, 27(2), 1978.

\bibitem{hitaj17}
B.~Hitaj, G.~Ateniese, and F.~Perez-Cruz.
\newblock Deep models under the gan: Information leakage from collaborative
  deep learning.
\newblock In {\em Proceedings of the 2017 ACM SIGSAC Conference on Computer and
  Communications Security}, CCS '17, pages 603--618, New York, NY, USA, 2017.
  ACM.

\bibitem{hu22}
H.~Hu, Z.~Salcic, L.~Sun, G.~Dobbie, P.~S. Yu, and X.~Zhang.
\newblock Membership inference attacks on machine learning: A survey.
\newblock {\em ACM Computing Surveys (CSUR)}, 54(11s):1--37, 2022.

\bibitem{huang20}
Y.~Huang, Z.~Song, K.~Li, and S.~Arora.
\newblock {I}nsta{H}ide: Instance-hiding schemes for private distributed
  learning.
\newblock In H.~D. III and A.~Singh, editors, {\em Proceedings of the 37th
  International Conference on Machine Learning}, volume 119 of {\em Proceedings
  of Machine Learning Research}, pages 4507--4518. PMLR, 13--18 Jul 2020.

\bibitem{huang22}
Z.~Huang, W.~jie Lu, C.~Hong, and J.~Ding.
\newblock Cheetah: Lean and fast secure {Two-Party} deep neural network
  inference.
\newblock In {\em 31st USENIX Security Symposium (USENIX Security 22)}, pages
  809--826, Boston, MA, Aug. 2022. USENIX Association.

\bibitem{jagielski2020}
M.~Jagielski, N.~Carlini, D.~Berthelot, A.~Kurakin, and N.~Papernot.
\newblock High accuracy and high fidelity extraction of neural networks.
\newblock In {\em 29th USENIX security symposium (USENIX Security 20)}, pages
  1345--1362, 2020.

\bibitem{kairouz19}
P.~Kairouz and Others.
\newblock Advances and open problems in federated learning.
\newblock {\em Foundations and Trends in Machine Learning}, 14(1-2), 2021.

\bibitem{katz07}
J.~Katz and Y.~Lindell.
\newblock {\em Introduction to Modern Cryptography}.
\newblock Chapman and Hall/CRC, 2007.

\bibitem{krizhevsky17}
A.~Krizhevsky, I.~Sutskever, and G.~E. Hinton.
\newblock Imagenet classification with deep convolutional neural networks.
\newblock {\em Commun. ACM}, 60(6):84--90, May 2017.

\bibitem{lfw}
G.~B. H.~E. Learned-Miller.
\newblock Labeled faces in the wild: Updates and new reporting procedures.
\newblock Technical Report UM-CS-2014-003, University of Massachusetts,
  Amherst, May 2014.

\bibitem{lee96_morph}
S.~Lee, K.~Chwa, J.~Hahn, and S.~Shin.
\newblock Image morphing using deformation techniques.
\newblock {\em JOURNAL OF VISUALIZATION AND COMPUTER ANIMATION}, 7(1):3 -- 23,
  n.d.

\bibitem{li17}
M.~Li, L.~Lai, N.~Suda, V.~Chandra, and D.~Z. Pan.
\newblock Privynet: {A} flexible framework for privacy-preserving deep neural
  network training with {A} fine-grained privacy control.
\newblock {\em CoRR}, abs/1709.06161, 2017.

\bibitem{mansfield15}
S.~Mansfield-Devine.
\newblock The {A}shley {M}adison affair.
\newblock {\em Network Security}, 2015(9):8 -- 16, 2015.

\bibitem{mohassel17}
P.~Mohassel and Y.~Zhang.
\newblock {SecureML}: A system for scalable privacy-preserving machine
  learning.
\newblock In {\em 2017 IEEE Symposium on Security and Privacy (SP)}, pages
  19--38, 2017.

\bibitem{ng21}
L.~K.~L. Ng, S.~S.~M. Chow, A.~P.~Y. Woo, D.~P.~H. Wong, and Y.~Zhao.
\newblock Goten: Gpu-outsourcing trusted execution of neural network training.
\newblock {\em Proceedings of the AAAI Conference on Artificial Intelligence},
  35(17):14876--14883, May 2021.

\bibitem{raff19}
E.~{Raff}, J.~{Sylvester}, S.~{Forsyth}, and M.~{McLean}.
\newblock Barrage of random transforms for adversarially robust defense.
\newblock In {\em 2019 IEEE/CVF Conference on Computer Vision and Pattern
  Recognition (CVPR)}, pages 6521--6530, 2019.

\bibitem{rathee20}
D.~Rathee, M.~Rathee, N.~Kumar, N.~Chandran, D.~Gupta, A.~Rastogi, and
  R.~Sharma.
\newblock Cryptflow2: Practical 2-party secure inference.
\newblock In {\em 27th Annual Conference on Computer and Communications
  Security (ACM CCS 2020)}. ACM, 2020.

\bibitem{sharma19}
S.~Sharma and K.~Chen.
\newblock Confidential boosting with random linear classifiers for outsourced
  user-generated data.
\newblock In {\em Computer Security - {ESORICS} 2019 - 24th European Symposium
  on Research in Computer Security, Luxembourg, September 23-27, 2019,
  Proceedings, Part {I}}, pages 41--65, 2019.

\bibitem{sharma18ic}
S.~Sharma, K.~Chen, and A.~Sheth.
\newblock Toward practical privacy-preserving analytics for iot and cloud-based
  healthcare systems.
\newblock {\em IEEE Internet Computing}, 22(2):42--51, Mar./Apr. 2018.

\bibitem{reza15}
R.~Shokri and V.~Shmatikov.
\newblock Privacy-preserving deep learning.
\newblock In {\em Proceedings of the 22nd ACM SIGSAC Conference on Computer and
  Communications Security}, 2015.

\bibitem{shokri15}
R.~Shokri and V.~Shmatikov.
\newblock Privacy-preserving deep learning.
\newblock In {\em Proceedings of the 22nd ACM SIGSAC Conference on Computer and
  Communications Security}, 2015.

\bibitem{shokri16}
R.~Shokri, M.~Stronati, C.~Song, and V.~Shmatikov.
\newblock Membership inference attacks against machine learning models.
\newblock In {\em 2017 {IEEE} Symposium on Security and Privacy, {SP} 2017, San
  Jose, CA, USA, May 22-26, 2017}, pages 3--18, 2017.

\bibitem{tramer18}
F.~Tramer and D.~Boneh.
\newblock Slalom: Fast, verifiable and private execution of neural networks in
  trusted hardware.
\newblock In {\em International Conference on Learning Representations}, 2019.

\bibitem{tramer16}
F.~Tram\`{e}r, F.~Zhang, A.~Juels, M.~K. Reiter, and T.~Ristenpart.
\newblock Stealing machine learning models via prediction apis.
\newblock In {\em Proceedings of the 25th USENIX Conference on Security
  Symposium}, SEC'16, pages 601--618, USA, 2016. USENIX Association.

\bibitem{unger15}
L.~Unger.
\newblock Breaches to customer account data.
\newblock {\em Computer and Internet Lawyer}, 32(2):14 -- 20, 2015.

\bibitem{maaten08}
L.~van~der Maaten and G.~Hinton.
\newblock Visualizing data using {t-SNE}.
\newblock {\em Journal of Machine Learning Research}, 9:2579--2605, 2008.

\bibitem{vempala05}
S.~S. Vempala.
\newblock {\em The Random Projection Method}.
\newblock American Mathematical Society, 2005.

\bibitem{xie14}
P.~Xie, M.~Bilenko, T.~Finley, R.~Gilad{-}Bachrach, K.~E. Lauter, and
  M.~Naehrig.
\newblock Crypto-nets: Neural networks over encrypted data.
\newblock {\em CoRR}, abs/1412.6181, 2014.

\bibitem{zhang18}
Q.-s. Zhang and S.-c. Zhu.
\newblock Visual interpretability for deep learning: a survey.
\newblock {\em Frontiers of Information Technology and Electronic Engineering},
  2018.

\bibitem{zhang20}
Y.~Zhang, R.~Jia, H.~Pei, W.~Wang, B.~Li, and D.~Song.
\newblock The secret revealer: Generative model-inversion attacks against deep
  neural networks.
\newblock In {\em CVPR}, 2020.

\end{thebibliography}
\end{document}